\documentclass[journal]{IEEEtran}
\usepackage{setspace}
\usepackage{amsmath}
\usepackage{amssymb}
\usepackage{amsfonts}
\usepackage{algorithm}
\usepackage{dsfont}
\usepackage{float}
\usepackage{cite}
\usepackage{graphicx}
\usepackage{epsfig}
\usepackage{subfigure}
\usepackage{psfrag}
\usepackage{xcolor}
\usepackage{url}
\usepackage{verbatim}
\usepackage{multirow}
\usepackage{subfigure}
\usepackage[colorlinks,linkcolor=black,urlcolor=black,anchorcolor=black,citecolor=black,hyperfootnotes=true]{hyperref}
\DeclareMathOperator*{\esssup}{ess\,sup}

\title{{Detecting Abrupt Change in Channel Covariance Matrix for MIMO Communication}
\thanks{Manuscript received July 5, 2022, revised December 9, 2022 and March 5, 2023, accepted March 7, 2023. The materials in this paper have been
presented in part at the IEEE Global Communications Conference (Globecom) 2021 \cite{9685287}. The work of R. Liu, D. He, and W. Zhang was supported by National Natural Science Foundation of China Program (62271316), National Key R\&D Project of China (2019YFB1802703), the Fundamental Research Funds for the Central Universities and Shanghai Key Laboratory of Digital Media Processing (STCSM 18DZ2270700). The work of L. Liu was supported by the Research Grants Council, Hong Kong SAR, China, under Grant 25215020. The work of E. G. Larsson was supported by ELLIIT and the KAW foundation. The associate editor coordinating the review of this paper and approving it for publication was Markku Juntti. (Corresponding author: Liang Liu.)}
\thanks{R. Liu, D. He and W. Zhang are with the Cooperative Medianet Innovation Center (CMIC), Shanghai Jiao Tong University, Shanghai, China (emails: {liurunnan, hedazhi, zhangwenjun}@sjtu.edu.cn). R. Liu is also with the Department of Electronic and Information Engineering, the Hong Kong Polytechnic University, Hong Kong SAR, China.}
\thanks{L. Liu is with the Department of Electronic and Information Engineering, The Hong Kong Polytechnic University, Hong Kong SAR, China (e-mail: liang-eie.liu@polyu.edu.hk).}
\thanks{E. G. Larsson is with the Department of Electrical Engineering, Link{\"o}ping University, 58183 Link{\"o}ping, Sweden (e-mail: erik.g.larsson@liu.se)}}
\author{\IEEEauthorblockN{Runnan Liu, \IEEEmembership{Graduate Student Member, IEEE}, Liang Liu, \IEEEmembership{Senior Member, IEEE}, Dazhi He, \IEEEmembership{Member, IEEE}, \\
		Wenjun Zhang, \IEEEmembership{Fellow, IEEE}, and Erik G. Larsson, \IEEEmembership{Fellow, IEEE}}
	\vspace{-0.7cm}}

\begin{document}
\maketitle \thispagestyle{empty} \vspace{-0.3in}

\newtheorem{thm}{\textbf{\textup{Theorem}}}
\newtheorem{Lem}{\textbf{\textup{Lemma}}}
\newtheorem{remark}{Remark}
\newtheorem{Prob}{Problem}
\newtheorem{Prop}{Proposition}
\newcommand{\tabincell}[2]{\begin{tabular}{@{}#1@{}}#2\end{tabular}}
\providecommand{\keywords}[1]{\textbf{\textit{Index terms---}} #1}

\begin{abstract}
The acquisition of the channel covariance matrix is of paramount importance to many strategies in multiple-input-multiple-output (MIMO) communications, such as the minimum mean-square error (MMSE) channel estimation. Therefore, plenty of efficient channel covariance matrix estimation schemes have been proposed in the literature. However, an abrupt change in the channel covariance matrix may happen occasionally in practice due to the change in the scattering environment and the user location. Our paper aims to adopt the classic change detection theory to detect the change in the channel covariance matrix as accurately and quickly as possible such that the new covariance matrix can be re-estimated in time. Specifically, this paper first considers the technique of on-line change detection (also known as quickest/sequential change detection), where we need to detect whether a change in the channel covariance matrix occurs at each channel coherence time interval. Next, because the complexity of detecting the change in a high-dimension covariance matrix at each coherence time interval is too high, we devise a low-complexity off-line strategy in massive MIMO systems, where change detection is merely performed at the last channel coherence time interval of a given time period. Numerical results show that our proposed on-line and off-line schemes can detect the channel covariance change with a small delay and a low false alarm rate. Therefore, our paper theoretically and numerically verifies the feasibility of detecting the channel covariance change accurately and quickly in practice.
\end{abstract}


\section{Introduction}
	\subsection{Motivation}
		Multiple-input-multiple-output (MIMO) is one of the core technologies in modern wireless communication systems to improve the network throughput, link range, and data reliability \cite{bolcskei2006space,gesbert2007shifting,goldsmith2003capacity,heath2016overview}.
		In the block fading environment where the MIMO channel is unknown at the beginning of each coherence time interval, it was shown in \cite{zheng2002communication} that at the high signal-to-noise ratio (SNR) regime, the MIMO channel capacity is achieved by first using the pilot signals to estimate the channels and then using the optimal precoding and decoding strategies for processing data signals based on the estimated channels. Motivated by this result, tremendous pioneering works have been done in the literature for estimating the instantaneous channel state information (CSI) \cite{Hassibi03,Gershman06} and optimizing the transmitter (Tx) precoding and receiver (Rx) decoding strategies with known CSI \cite{jorswieck2008complete,cadambe2008interference,zhang2010cooperative,wiesel2006linear,yu2004sum}.
				
		In practice, the MIMO channels are spatially correlated due to the dependent antenna patterns and the finite number of multipath components from the propagation environments \cite{Gao15,Sanguinetti20}.
		Note that besides the instantaneous CSI, the acquisition of the channel covariance matrix is essential in many scenarios of MIMO communication as well.
		For example, the minimum mean-squared error (MMSE) technique is widely used to estimate the instantaneous CSI \cite{Marzetta16,ngo2013energy,hoydis2013massive}. However, the MMSE channel estimator is a function of the channel covariance matrix.
		Moreover, pilot contamination due to pilot reuse in multiple cells and training overhead due to the large number of Tx antennas are the main challenges limiting the performance of the uplink and downlink channel estimation in massive MIMO systems, respectively. Recently, it was shown that if the channel covariance matrix can be known and properly utilized, the effect of uplink pilot contamination and downlink training overhead on network throughput can be significantly reduced \cite{yin2013coordinated,ngo2012evd,yin2013decontaminating,bjornson2017massive,6542746}.
		Furthermore, under the newly emerging intelligent reflecting surface (IRS) assisted MIMO communication, it is a challenge to estimate all the instantaneous channels associated with the huge number of IRS reflecting elements for designing the IRS reflecting coefficients. To reduce the channel estimation overhead, recently, plenty of works have utilized the channel covariance matrix, instead of the instantaneous CSI, to design the IRS reflecting elements for optimizing the long-term performance \cite{Park17,Jin19,Zhao21}.
		
		In the literature, many efficient estimation algorithms have been proposed for acquiring the channel covariance matrix in MIMO systems \cite{Chin,upadhya2018covariance}. It is worth noting that despite slowly, the channel covariance matrix does change in practice due to the mobility of users and the change in the scattering environment around users and/or the base stations (BSs) \cite{goldsmith2003capacity}.
		An open question that is not addressed in \cite{Chin,upadhya2018covariance} is that when an abrupt change in the channel covariance matrix occurs, how to detect it accurately and quickly so that we can re-estimate the new channel covariance matrix based on the algorithms proposed in these works as soon as possible. An intuitive way to tackle the above issue is to first estimate the channel covariance matrix very accurately and then check whether a change has occurred over the original channel covariance matrix. However, an accurate covariance estimation can only be obtained based on a sufficiently large number of estimated channels over a very long time. Hence, the delay to detect the covariance change based on the above method is too long. In practice, we hope that the change can be detected as quickly as possible such that the covariance-based communication strategies \cite{Marzetta16,ngo2013energy,hoydis2013massive,yin2013coordinated,ngo2012evd,yin2013decontaminating,bjornson2017massive,6542746,Park17,Jin19,Zhao21} can stop using the original covariance matrix immediately after a change occurs. Fortunately, detection is usually easier than estimation. For example, suppose that a random variable is uniformly distributed in the interval $[0,1]$ at the beginning. If we observe a sample of $2$ at a moment, then we can confidently declare a distribution change based on this single sample, even if we do not know the exact distribution after change. Motivated by this potential, in this paper, we aim to adopt the classic change detection theory \cite{poor2008quickest,tartakovsky2014sequential} to detect the change in the channel covariance matrix accurately and quickly.
		
		\subsection{Prior Work}
		The technique of change detection based on stochastic observations has a wide range of applications for quality control \cite{goldsmith1961average}, navigation system monitoring \cite{4107517}, seismic data processing \cite{kitagawa1985smoothness}, segmentation of signals \cite{andre1988new}, etc. Typically, the change detection methods can be classified into on-line and off-line ones.
		
		First, the on-line change detection, also referred to as the sequential change detection or the quickest change detection, aims to detect the change as quickly as possible, subject to a false alarm constraint. Specifically, let $u_k$ denote the sample observed at time instant $k~(k\geq 1)$, with a conditional probability density function (pdf) $p_\phi(u_k|u_1,\dots,u_{k-1})$ \footnote{If $u_k$'s are independent with each other, then the conditional pdf of $u_k$ reduces to $p_\phi(u_k)$.}, where $\phi$ denotes the pdf parameter, e.g., mean, variance, etc. Let us assume that $\phi=\phi_0$ before the change occurs and $\phi=\phi_1$ after the change occurs. Then, under the on-line change detection framework, we keep calculating the value of  some confidence function  $g_k\big(\phi_0,\phi_1,u_1,\dots,u_k\big)$ at each time instant $k$, which is a properly designed function of the samples observed until the current instant and the pdf parameters $\phi_0, \phi_1$. If at a time instant $k$, the value of the confidence function is above a threshold $\theta$, i.e., $g_k\big(\phi_0,\phi_1,u_1,\dots,u_k\big)>\theta$, we declare that a change in data distribution is detected.
		
		Second, under the off-line change detection framework, we merely check whether a change in distribution occurs after we have all the observations over a period of $K$ time instants, i.e., $u_1,\dots,u_K$. Specifically, we declare that the distribution has changed at some moment if $g_K\big(\phi_0,\phi_1,u_1,\dots,u_K\big)>\theta$ and has not changed otherwise. Since the on-line scheme can check whether a change occurs at each time instant, its detection delay is shorter than that of the off-line counterpart. However, the off-line scheme is of lower computational complexity, which is appealing to applications with high-dimension data sets.
		
		Lastly, for both on-line and off-line schemes, a good change detector should have a good trade-off between the ability to detect the change quickly when a change occurs, i.e., the detection delay performance, and the ability to not detect the change when no change occurs, i.e., the false alarm constraint. To achieve this goal, in the literature, the confidence functions $g_k\big(\phi_0,\phi_1,u_1,\dots,u_k\big)$'s are usually designed based on the log-likelihood ratios (LLRs) between $p_{\phi_0}(u_k|u_1,\dots,u_{k-1})$'s and $p_{\phi_1}(u_k|u_1,\dots,u_{k-1})$'s, which can minimize the detection delay subject to the false alarm constraint according to Neyman-Pearson Lemma \cite{neyman1933ix}. Moreover, other signal processing technique, e.g., cumulative score statistic \cite{box1992cumulative} and generalized score statistic \cite{liu2021score}, were also utilized to optimize the above trade-off in the change detection.
		
		In the literature, several works have considered the application of change detection theory in wireless communication. Specifically, the on-line change detection technique was applied in cognitive network to detect the moment when the primary user changes from the active state to the inactive state such that the secondary user can start to transmit when the primary user becomes inactive \cite{4698342}.
		More relevant to our work, \cite{9014219} and \cite{9500529} studied the on-line change detection schemes to identify the change in the millimeter wave (mmWave) MIMO channel caused by the movement of the obstacles, scatters, etc. However, these change detectors are not designed based on the LLR criterion and not optimal for the trade-off between detection delay and false alarm.
		In this paper, we will design both the on-line and off-line schemes to detect the change in the MIMO channel covariance matrix based on the LLR criterion, which is new in the literature to our best knowledge.
		
		\subsection{Main Contributions}
		In this paper, we consider the point-to-pint MIMO communication in a rich scattering environment, where the MIMO channel follows the Rayleigh fading model. Hence, the channel distribution merely depends on the covariance matrix, which may change abruptly at some unknown moment. Our goal is to detect the change in the channel covariance matrix accurately and quickly such that the new channel covariance matrix after change can be estimated in due time for benefiting the covariance-based communication strategies \cite{Marzetta16,ngo2013energy,hoydis2013massive,yin2013coordinated,ngo2012evd,yin2013decontaminating,bjornson2017massive,6542746,Park17,Jin19,Zhao21}. The contributions of this paper are summarized as follows.
		
		First, we consider the genie-aided on-line change detection scheme, where the channel covariance matrix after change is assume to be known ahead of time. In particular, the cumulative-sum (CUSUM) algorithm \cite{page1954continuous}, where the confidence function $g_k\big(\phi_0,\phi_0,u_1,\dots,u_k\big)$ at time instant $k$ is defined as the sum of LLRs until time instant $k$, is utilized for change detection. Moreover, under the CUSUM framework, we also analytically characterize the trade-off between the false alarm rate (FAR) and the conditional average detection delay (CADD) \cite{pollak1985optimal}. These results provide lots of insights on the design and analysis of the on-line change detectors.
		
		Second, we consider the on-line change detection in the practical case when the channel covariance matrix after change is unknown. With unknown channel covariance, at each time instant $k$, the CUSUM technique requires a rough estimation of the channel covariance matrix based on the estimated channels from time instant $1$ to time instant $k$, an estimation based on the estimated channel from time instant $2$ to time instant $k$, and so on. To reduce the computational complexity, in this paper, we consider a window-limited generalized-likelihood-ratio (WL-GLR) framework \cite{willsky1976generalized}, where at each time instant, the new channel covariance matrix is roughly estimated by a very small number of times based on the maximum likelihood (ML) technique \cite{aubry2012maximum}. Numerical results show that the WL-GLR strategy can achieve very good detection performance with significantly reduced complexity.
		
		Lastly, we consider the off-line scheme in massive MIMO systems which merely detects whether the channel covariance matrix has changed once at the last time instant of a period of time.
		The reason to advocate the off-line scheme here is that the dimension of the channel covariance matrix in massive MIMO systems is very high and it is practically difficult to estimate it several times at each time instant as required by the on-line change detection counterpart.
		Moreover, instead of using the ML estimator, we adopt a shrinkage-based algorithm \cite{5484583} to roughly estimate the channel covariance matrix in massive MIMO systems for change detection, because such a scheme is known to generate a well-conditioned estimation even when the number of observations is much smaller than the dimension of the covariance matrix.
		
		\subsection{Organization}
		The rest of this paper is organized as follows. Section \ref{Sec_SysMod} describes the system model for channel covariance matrix change detection. Section \ref{Protocols} introduces the proposed on-line and off-line channel covariance matrix change detection protocols. Section \ref{Sec_KnownC} and \ref{Sec_unKnownC} investigate the on-line change detection with known and unknown channel covariance matrix after change, respectively. The off-line change detection scheme for massive MIMO systems is designed in Section \ref{SecIII}. Numerical results are provided in Section \ref{NumRes}. Finally, Section \ref{Conclusion} concludes the paper and outlines the future research directions.
		
		\textit{Notations:} In this paper, letter $\boldsymbol{a}$, $\boldsymbol{A}$, and $\mathcal{A}$ denote vector, matrix and set, respectively. In addition,
		$(\boldsymbol{A})^H$, $|\boldsymbol{A}|$, and $\text{tr}(\boldsymbol{A})$ denote conjugate transpose, determinant and trace of $\boldsymbol{A}$, respectively. Further, $\text{vec}(\boldsymbol{A})$ denotes the column vector of $\boldsymbol{A}$. The notation $\text{diag}(\boldsymbol{a})$ is used for a diagonal matrix whose diagonal elements are the elements of $\boldsymbol{a}$ in the corresponding position. Moreover, $\text{diag}(\boldsymbol{A})$ stands for a vector whose element is the diagonal element of matrix $\boldsymbol{A}$. Operator $\text{mod}(\cdot)$ denotes a modulo operation. Operator $\lfloor\cdot\rfloor$ denotes the floor operator which outputs an integer smaller or equal to its input value.

\section{System Model}\label{Sec_SysMod}

Consider a narrow-band MIMO communication system where the Tx and the Rx are equipped with $M_t$ and $M_r$ antennas, respectively. The channel between the Tx and the Rx is denoted by  $\boldsymbol{H}\in \mathbb{C}^{M_t\times M_r}$. We assume a Rayleigh fading channel with $\boldsymbol{h}=\text{vec}(\boldsymbol{H})\sim \mathcal{CN}(\boldsymbol{0},\boldsymbol{C})$, where $\boldsymbol{C}\in \mathbb{C}^{M\times M}\succ \boldsymbol{0}$ denotes the channel covariance matrix with $M=M_tM_r$. Moreover, consider a block fading channel model, where $\boldsymbol{H}$ stays constant in one coherence time interval, but may vary independently over different intervals.  In practice, the channel covariance matrix $\boldsymbol{C}$ usually changes much more slowly over time as compared to the instantaneous channel $\boldsymbol{H}$. For all the communication protocols relying on the knowledge of the channel covariance matrix, e.g., MMSE channel estimation, any notable change in $\boldsymbol{C}$ should be detected in an accurate and timely manner, such that the system can re-estimate the new channel covariance matrix immediately for its usage. In this paper, we adopt the change detection technique to achieve the above goal.

Specifically, since the channel covariance matrix changes at a larger timescale as compared to the channel coherence interval, in this paper, we define a covariance interval as a sequence of channel coherence intervals over which a change in the channel covariance matrix may occur. Let $L_i$ denote the number of coherence time intervals contained in the $i$-th covariance interval.
For convenience, we further define $\boldsymbol{H}_{i,j}$ as the channel at the $j$-th coherence time interval of the $i$-th covariance interval and $\boldsymbol{C}_{i,j}$ as the covariance matrix of $\boldsymbol{h}_{i,j} = \text{vec}(\boldsymbol{H}_{i,j})$, $j=1,\dots,L_i,~\forall i$.
As a result, if no change in channel covariance occurs in covariance interval $i$, then we have
	\begin{equation}
		\boldsymbol{C}_{i,j} = \tilde{\boldsymbol{C}}_{i-1}, ~j=1,\dots,L_i,
	\end{equation}
	where $\tilde{\boldsymbol{C}}_{i-1}=\boldsymbol{C}_{i-1,L_{i-1}}$ is the channel covariance matrix at the end of covariance interval $i-1$. Otherwise, if a change in channel covariance occurs in coherence time interval $\nu_i$ of covariance interval $i$, then we have
	\begin{equation}
		\boldsymbol{C}_{i,j}=\left\{
		\begin{aligned}
			&\tilde{\boldsymbol{C}}_{i-1},~j=1,\dots,\nu_i-1,\\
			&\tilde{\boldsymbol{C}}_i,~~~j=\nu_i,\dots,L_i,
		\end{aligned}
		\right.
	\end{equation}
	where $\tilde{\boldsymbol{C}}_i$ is the new channel covariance matrix after change in covariance interval $i$. As a result, at the first a few coherence time intervals of each covariance interval, we need to check whether the channel covariance matrix has changed at some time.
	Furthermore, if a change is detected in a covariance interval, we need to spend some other coherence time intervals in this covariance interval to estimate the new channel covariance matrix.

\begin{figure}[t]
	\centering
	\includegraphics[width=9cm]{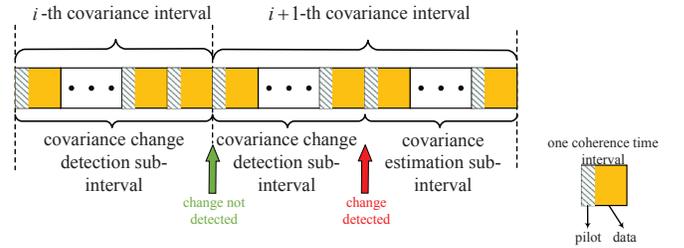}
	\vspace{-0.3cm}
	\caption{Communication protocol with channel covariance matrix change detection and estimation taken into consideration.}
	\label{Framework_fig}
	\vspace{-0.5cm}
\end{figure}
	
	In practice, a communication framework with channel covariance matrix change detection and the corresponding estimation taken into consideration can work as follows. As shown in Fig. \ref{Framework_fig}, each covariance interval can be divided into two sub-intervals, i.e., the covariance change detection sub-interval and the covariance estimation sub-interval (which is optional as shown later in the paper). Let $L_{i,1}>0$ and $L_{i,2}\geq0$ denote the numbers of coherence time intervals in the first and the second sub-intervals of covariance interval $i$, respectively, such that $L_i=L_{i,1}+L_{i,2},~\forall i$. In the first sub-interval of a covariance interval $i$, the pilot signal received at the $j$-th coherence time interval is
	\begin{equation}
	 	\begin{aligned}
	 		\boldsymbol{Y}_{i,j} = \sqrt{\rho}\boldsymbol{X}\boldsymbol{H}_{i,j}+\boldsymbol{N}_{i,j},~j\leq L_{i,1},
	 	\end{aligned}
	 	\label{Eq_SysRxSignal}
	 \end{equation}
	where $\rho$ denotes the transmit power, $\boldsymbol{X}\in \mathbb{C}^{T\times M_t}$ with $\text{tr}(\boldsymbol{X}^H\boldsymbol{X}) = T$ denotes the pilot sequence consisting of $T$ symbols, and $\boldsymbol{N}_{i,j}$ denotes the additive white Gaussian noise (AWGN) at the Rx side with $\textup{vec}(\boldsymbol{N}_{i,j})\sim \mathcal{CN}(\boldsymbol{0},\sigma^2 \boldsymbol{I})$. In covariance interval $i$, based on all the pilot signals received in the first sub-interval, i.e., $\boldsymbol{Y}_{i,j},~j=1,\dots, L_{i,1}$, the change detection job is to detect between the following two hypothesis
	\begin{equation}
		\begin{aligned}
			H_0:~& \boldsymbol{C}_{i,j}=\widetilde{\boldsymbol{C}}_{i-1},~~~~~~~~~~1\leq j\leq L_{i,1},\\
			H_1:~& \exists ~\nu_i<L_{i,1}, ~\text{such that:}\\
			& \boldsymbol{C}_{i,j}=\widetilde{\boldsymbol{C}}_{i-1},~~~~~~~~~~1\leq j< \nu_i,\\
			& \boldsymbol{C}_{i,j}=\widetilde{\boldsymbol{C}}_{i}\neq\widetilde{\boldsymbol{C}}_{i-1},~~~\nu_i\leq j\leq L_{i,1}.
		\end{aligned}
		\label{Hypos}
	\end{equation}
	If no change is detected in the first sub-interval of covariance interval $i$, i.e., $H_0$ is true, then the channel covariance matrix estimated in the last covariance interval, i.e., $\tilde{\boldsymbol{C}}_{i-1}$, can still be used by the system, i.e., $\widetilde{\boldsymbol{C}}_i=\widetilde{\boldsymbol{C}}_{i-1}$. As a result, we can skip the second sub-interval for covariance estimation, i.e., $L_{i,2}=0$, and start a new covariance interval $i+1$ for change detection. Otherwise, if a change is detected in the first sub-interval of covariance interval $i$, i.e., $H_1$ is true, then the new channel covariance matrix, i.e., $\tilde{\boldsymbol{C}}_i\neq \widetilde{\boldsymbol{C}}_{i-1}$, should be estimated based on the pilot signals received in the second sub-interval. After the new channel covariance matrix is estimated, we can start a new covariance interval $i+1$ for change detection.
	
	Since the covariance matrix estimation has been widely studied in the literature, e.g.,  \cite{Chin} and \cite{upadhya2018covariance}, in the following, we focus on the proposed change detection approach to detect whether the channel covariance matrix has changed at some coherence time interval based on the pilot signals received in the first sub-interval of a covariance interval given in (\ref{Eq_SysRxSignal}).
	
	\section{Channel Covariance Matrix Change Detection Protocols}\label{Protocols}
	In the first sub-interval of a covariance interval $i$, the received pilot signals are given in (\ref{Eq_SysRxSignal}). In this paper, we assume that $T\geq M_t$ such that the pilot sequences of different transmit antennas can be orthogonal to each other, i.e., $\boldsymbol{X}^H \boldsymbol{X}= \frac{T}{M_t}\boldsymbol{I}$.
	Note that before change detection is performed at interval $i$, we are not sure about whether the channel covariance matrix remains to be $\tilde{\boldsymbol{C}}_{i-1}$ at all the coherence time intervals or not, As a result, we adopt the ML technique to estimate the channel at each coherence time interval, which does not rely on the channel statistics, for change detection in the first sub-interval of each covariance interval.\footnote{However, all the other covariance-based communication protocols, e.g., MMSE channel estimation for data decoding, can still use $\tilde{\boldsymbol{C}}_{i-1}$ at the beginning of interval $i$ until a change is detected at some coherence time interval. Since our proposed scheme can detect the change quickly, the outdated covariance matrix will not be used for a long time in these protocols} Specifically, at the $j$-th coherence time interval of the first sub-interval of covariance interval $i$, the ML channel estimator is given by
	\begin{equation}
		\begin{aligned}
			\bar{\boldsymbol{H}}_{i,j} =\frac{M_t}{\sqrt{\rho}T}\boldsymbol{X}^H\boldsymbol{Y}_{i,j}=\boldsymbol{H}_{i,j} + \bar{\boldsymbol{N}}_{i,j}, ~~j = 1,\dots,L_{i,1},
		\end{aligned}
		\label{tildey}
	\end{equation}
	where $\bar{\boldsymbol{N}}_{i,j}=\frac{M_t}{\sqrt{\rho} T}\boldsymbol{X}^H\boldsymbol{N}_{i,j}$ with $\textup{vec}(\bar{\boldsymbol{N}}_{i,j})\sim \mathcal{CN}(\boldsymbol{0},\frac{\sigma^2}{E_0}\boldsymbol{I})$ and $E_0=\frac{\rho T}{M_t^2}$.
	
	One intuitive way to perform change detection is to first estimate the channel covariance matrix in each covariance interval based on the estimated channels $\bar{\boldsymbol{H}}_{i,j}$'s very accurately and then detect whether a significant change in the channel covariance matrix occurs. However, an accurate estimation of the channel covariance matrix requires a large number of estimated channels in (\ref{tildey}), which is not suitable for detecting the change with a small delay.
	According to the change detection theory \cite{basseville1993detection}, a change can be detected quickly even without a very accurate estimation of the new distribution. Following the above change detection theory, in this paper, we will propose a CUSUM-based approach to detect the change in the channel covariance matrix quickly and accurately.
	
	First, we define $\bar{\boldsymbol{h}}_{i,j}=\text{vec}(\bar{\boldsymbol{H}}_{i,j})\in\mathbb{C}^{M\times 1},~j=1,$ $\dots,L_{i,1},~\forall i$. According to (\ref{tildey}), we have $\bar{\boldsymbol{h}}_{i,j}\in\mathcal{CN}(\boldsymbol{0},\boldsymbol{C}_{i,j}+\frac{\sigma^2}{E_0}\boldsymbol{I}),~j=1,\dots,L_{i,1},~\forall i$. As a result, given any covariance matrix $\boldsymbol{C}_{i,j}$, the conditional PDF of $\bar{\boldsymbol{h}}_{i,j}$ is
	\begin{equation}
		\begin{aligned}
			p\big(\bar{\boldsymbol{h}}_{i,j}|\boldsymbol{C}_{i,j}\big) =& \frac{\exp\bigg[-(\bar{\boldsymbol{h}}_{i,j})^H\big(\boldsymbol{C}_{i,j}+\frac{\sigma^2}{E_0}\boldsymbol{I}\big)^{-1}\bar{\boldsymbol{h}}_{i,j}\bigg]}{\pi^M\big|\boldsymbol{C}_{i,j}+\frac{\sigma^2}{E_0}\boldsymbol{I}\big|},\\
			&j=1,\dots,L_{i,1},~\forall i.
		\end{aligned}
		\label{Prob_i}
	\end{equation}
	Then, at the $j$-th coherence time interval of the $i$-th covariance interval, we define the LLR between the event of $\boldsymbol{C}_{i,j}=\tilde{\boldsymbol{C}}_i\neq\tilde{\boldsymbol{C}}_{i-1}$ and that of $\boldsymbol{C}_{i,j}=\tilde{\boldsymbol{C}}_{i-1}$ as
	\begin{equation}
		\begin{aligned}
			&LLR_{i,j}(\bar{\boldsymbol{h}}_{i,j},\widetilde{\boldsymbol{C}}_{i-1},\widetilde{\boldsymbol{C}}_i)\\
			\triangleq&\log\Big(p\big(\bar{\boldsymbol{h}}_{i,j}|\widetilde{\boldsymbol{C}}_i\big)\Big)-\log\Big(p\big(\bar{\boldsymbol{h}}_{i,j}|\widetilde{\boldsymbol{C}}_{i-1}\big)\Big), ~1\leq j\leq L_{i,1}, ~\forall i.
		\end{aligned}
		\label{LLR}
	\end{equation}
	
	Based on the LLRs available at the first sub-interval of the $i$-th covariance interval, we need to determine whether the channel covariance matrix remains to be $\tilde{\boldsymbol{C}}_{i-1}$, i.e., event $H_0$ is true, or has changed to some new covariance matrix $\tilde{\boldsymbol{C}}_{i}\neq\tilde{\boldsymbol{C}}_{i-1}$, i.e., event $H_1$ is true. In the rest of this section, we first propose an on-line protocol where the change detection is performed at \textit{each} of the coherence time intervals in the first sub-interval of a covariance interval, based on the LLRs. To reduce the implementation complexity arising from applying change detection at each coherence time interval, especially in massive MIMO systems where the dimension of data is extremely high, we then propose an off-line protocol, where at each covariance interval $i$, change detection is only performed at the last coherence time interval of the first sub-interval, i.e., coherence time interval $L_{i,1}$, based on the LLRs. In the following, we introduce these two change detection protocols in details.
	
	\begin{figure}[t]
		\vspace{-0.2cm}
		\centering
		\includegraphics[width=9cm]{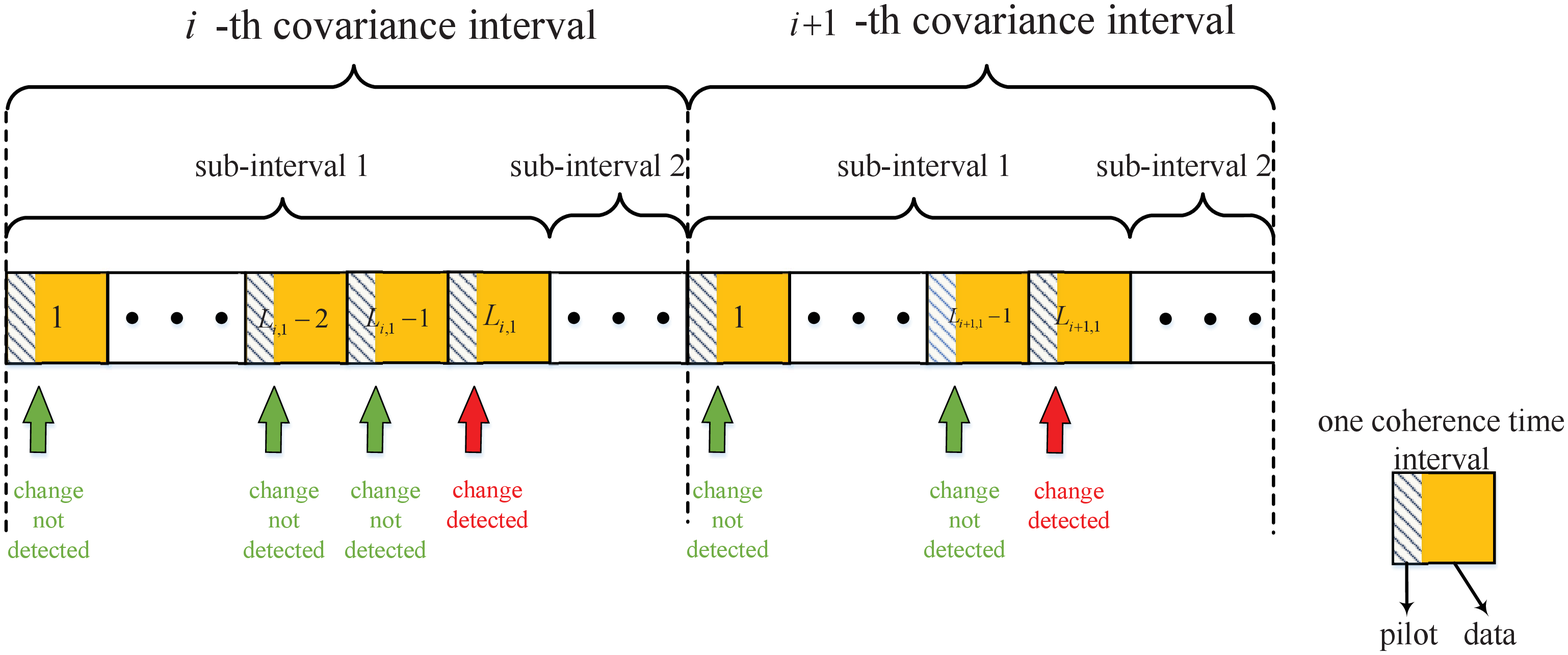}
		\vspace{-0.3cm}
		\caption{On-line change detection protocol.}
		\label{Online_fig}
		\vspace{-0.35cm}
	\end{figure}
	\textbf{On-line Change Detection Protocol for General MIMO}: As shown in Fig. \ref{Online_fig}, in the first sub-interval of a covariance interval $i$, we perform change detection in each coherence time interval. Specifically, at the $j$-th coherence time interval, $j \geq 1$, the change detection is performed based on all the LLRs obtained at and before this coherence time interval, i.e.,  $\big\{LLR_{i,1}(\bar{\boldsymbol{h}}_{i,1},\widetilde{\boldsymbol{C}}_{i-1},\widetilde{\boldsymbol{C}}_i),\dots,$ $LLR_{i,j}(\bar{\boldsymbol{h}}_{i,j},\widetilde{\boldsymbol{C}}_{i-1},\widetilde{\boldsymbol{C}}_i)\big\}$. If a change has been detected at some coherence time interval $j$, then the first sub-interval of covariance interval $i$ for change detection will end, i.e., $L_{i,1}=j$. Note that the length of the first sub-interval in different covariance intervals may vary because the change may be detected at different moments in different covariance intervals. Moreover, the second sub-interval for covariance matrix estimation is necessary under the on-line protocol after a change is detected.
	
\begin{figure}[t]
		\vspace{-0.2cm}
		\centering
		\includegraphics[width=8cm]{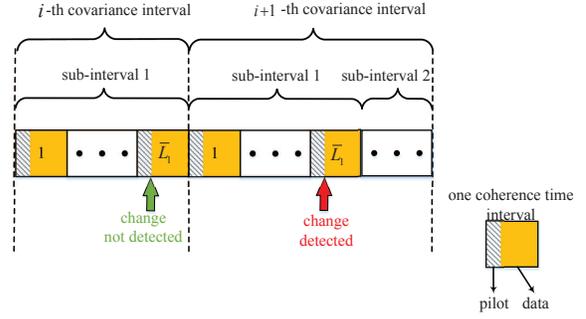}
		\vspace{-0.3cm}
		\caption{Off-line change detection protocol.}
		\label{Offline_fig}
		\vspace{-0.4cm}
	\end{figure}

	\textbf{Off-line Change Detection Protocol for Massive MIMO}: As shown in Fig. \ref{Offline_fig}, under this protocol, the numbers of coherence time intervals in the first sub-interval of all the covariance intervals for change detection are set to be identical, i.e., $L_{i,1}=\bar{L}_1,~\forall i$. Then, for each covariance interval $i$, the change detection is merely conducted at coherence time interval $\bar{L}_1$ based on all the LLRs obtained at and before this coherence time interval, i.e., $\big\{LLR_{i,1}(\bar{\boldsymbol{h}}_{i,1},\widetilde{\boldsymbol{C}}_{i-1},\widetilde{\boldsymbol{C}}_i),\dots,$ $LLR_{i,\bar{L}_1}(\bar{\boldsymbol{h}}_{i,\bar{L}_1},\widetilde{\boldsymbol{C}}_{i-1},\widetilde{\boldsymbol{C}}_i)\big\}$.
	Note that if $\bar{L}_1$ is too small, then we do not have sufficient observations to make accurate change detection.
	On the other hand, if $\bar{L}_1$ is too large, then the detection delay is too long. Moreover, multiple changes may occur in one or some consecutive covariance intervals, which makes change detection more challenging.
	Therefore, $\bar{L}_1$ should be carefully determined in practice to balance between detection accuracy and detection delay. Note that the second sub-interval for covariance estimation is optional under the off-line protocol, since it is likely that no change is declared at the end of the first sub-interval (please see the $i$-th covariance interval in Fig. \ref{Offline_fig}).
		
	In the rest of this paper, we will introduce how to design the change detectors under the on-line and off-line protocols, respectively.

	\section{On-line Change Detection with Known $\{\tilde{\boldsymbol{C}}_i\}$}\label{Sec_KnownC}
	In this section, we will devise the genie-aided on-line channel covariance matrix change detectors, where before the beginning of each covariance interval $i$, the new covariance matrix $\widetilde{\boldsymbol{C}}_i$ after a change occurs  is assumed to be known with the help of some genie. The results on this ideal case can help us get more insights on the on-line change detection. We will focus on the on-line change detection with unknown new covariance matrices in the next section.
	
	\subsection{CUSUM-based On-line Change Detectors}\label{SecIII-A}
	 If a change in the channel covariance matrix occurs in coherence time interval $\nu_i$ of covariance interval $i$, then at each coherence time interval $j<\nu_i$ ($j>\nu_i$), we have $LLR_{i,j}(\bar{\boldsymbol{h}}_{i,j},\widetilde{\boldsymbol{C}}_{i-1},\widetilde{\boldsymbol{C}}_i)<0$ ($LLR_{i,j}(\bar{\boldsymbol{h}}_{i,j},\widetilde{\boldsymbol{C}}_{i-1},\widetilde{\boldsymbol{C}}_i)>0$) with a very high probability, especially when $\tilde{\boldsymbol{C}}_i$ changes significantly over $\tilde{\boldsymbol{C}}_{i-1}$.
	 However, there is always a chance that the random channel $\bar{\boldsymbol{h}}_{i,j}$ at some coherence interval $j<\nu_i$ ($j>\nu_i$) makes $LLR_{i,j}(\bar{\boldsymbol{h}}_{i,j},\widetilde{\boldsymbol{C}}_{i-1},\widetilde{\boldsymbol{C}}_i)>0$ ($LLR_{i,j}(\bar{\boldsymbol{h}}_{i,j},\widetilde{\boldsymbol{C}}_{i-1},\widetilde{\boldsymbol{C}}_i)<0$). Therefore, instead of checking whether the LLR is positive or negative at one single moment, it is more reliable and robust to detect the change in the channel covariance matrix based on multiple LLRs over a sequence of coherence time intervals, i.e., a change will be declared if most of the LLRs in the past a few coherence time intervals are positive.
	
\begin{figure}[t]
		\vspace{-0.3cm}
		\centering
		\includegraphics[width=8.5 cm]{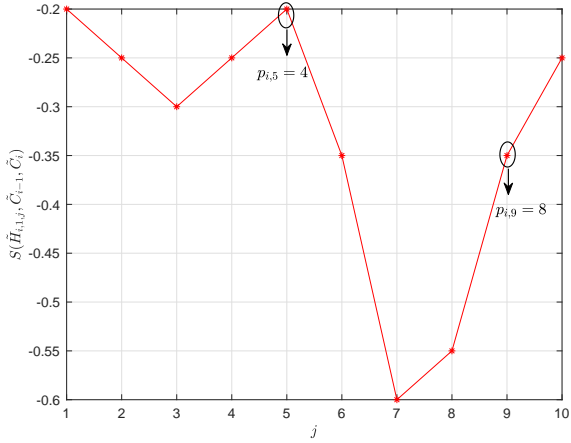}
		\caption{An illustration of $S(\tilde{\boldsymbol{H}}_{i,1,j},\widetilde{\boldsymbol{C}}_{i-1},\widetilde{\boldsymbol{C}}_i)$ with $j=1$ to $j=10$. It is assumed that $LLR_{i,j}(\bar{\boldsymbol{h}}_{i,j},\widetilde{\boldsymbol{C}}_{i-1},\widetilde{\boldsymbol{C}}_i)=-0.20,-0.05,-0.05$ $,0.05,0.05,-0.15,-0.25,0.05,0.20,0.10$ when $j$ ranges from $1$ to $10$. In this case, it can be seen that there are two transition points, $j=4$ and $j=8$. Moreover, $p_{i,j}$ also changes with $j$, e.g., $p_{i,5}=4$, but $p_{i,9}=8$.}
		\label{Schematic}
		\vspace{-0.4cm}
	\end{figure}

	To achieve the above goal, at the $j$-th coherence time interval of the first sub-interval of the $i$-th covariance interval, we first define the sum of LLRs from the $j_1$-th coherence time interval to the $j_2$-th coherence time interval of covariance interval $i$ as
	\begin{equation}
		\begin{aligned}
			S(\tilde{\boldsymbol{H}}_{i,j_1,j_2},\widetilde{\boldsymbol{C}}_{i-1},\widetilde{\boldsymbol{C}}_i)=&\sum_{j=j_1}^{j_2}LLR_{i,j}(\bar{\boldsymbol{h}}_{i,j},\widetilde{\boldsymbol{C}}_{i-1},\widetilde{\boldsymbol{C}}_i),\\
			&1\leq j_1\leq j_2\leq L_{i,1},
		\end{aligned}
		\label{S_on-line}
	\end{equation}
	where $\tilde{\boldsymbol{H}}_{i,j_1,j_2}=[\bar{\boldsymbol{h}}_{i,j_1},\cdots,\bar{\boldsymbol{h}}_{i,j_2}]$.
	Then, at the $j$-th coherence time interval of the first sub-interval of the $i$-th covariance interval, we define
	\begin{equation}
		\begin{aligned}
			&p_{i,j} = \arg\max_{1\leq p\leq j}S(\tilde{\boldsymbol{H}}_{i,p,j},\tilde{\boldsymbol{C}}_{i-1},\tilde{\boldsymbol{C}}_i),\\
			&W_{i,j}(\tilde{\boldsymbol{C}}_{i-1},\tilde{\boldsymbol{C}}_i) = S(\tilde{\boldsymbol{H}}_{i,p_{i,j},j},\tilde{\boldsymbol{C}}_{i-1},\tilde{\boldsymbol{C}}_i).
			\label{on-line_W}
		\end{aligned}
	\end{equation}
	Therefore, at the $i$-th covariance interval, for all the summations of the LLRs starting from the $p$-th coherence time interval ($1\leq p\leq j$) to the $j$-th coherence time interval, the one starting from the $p_{i,j}$-th coherence time interval achieves the maximum. This maximum is defined as $W_{i,j}(\tilde{\boldsymbol{C}}_{i-1},\tilde{\boldsymbol{C}}_i)$. Note that $p_{i,j}$ may change over $j$ (please see Fig. \ref{Schematic}). As a result, $p_{i,j}$ and $W_{i,j}(\tilde{\boldsymbol{C}}_{i-1},\tilde{\boldsymbol{C}}_i)$ should be calculated at each coherence time interval in the change detection phase. Note that according to \cite{basseville1993detection}, $W_{i,j}(\tilde{\boldsymbol{C}}_{i-1},\tilde{\boldsymbol{C}}_i)$'s, $\forall i, j$, can be calculated recursively, shown as follows
	\begin{equation}
		\begin{aligned}
			W_{i,j}(\tilde{\boldsymbol{C}}_{i-1},\tilde{\boldsymbol{C}}_i)&\\
			=\max\Big(W_{i,j-1}(&\tilde{\boldsymbol{C}}_{i-1},\tilde{\boldsymbol{C}}_i)+LLR_{i,j}(\bar{\boldsymbol{h}}_{i,j},\widetilde{\boldsymbol{C}}_{i-1},\widetilde{\boldsymbol{C}}_i),0\Big),\\
			& j\geq1,\forall i,
		\end{aligned}	
		\label{RecursiveForm}
	\end{equation}
	where $\max(a,0)$ denotes the maximum value in $a$ and $0$, and $W_{i,0}(\tilde{\boldsymbol{C}}_{i-1},\tilde{\boldsymbol{C}}_i)=0$. The above method can significantly reduce the complexity to update $W_{i,j}(\tilde{\boldsymbol{C}}_{i-1},\tilde{\boldsymbol{C}}_i)$'s.

Then, at the $j$-th coherence time interval of the $i$-th covariance interval, the CUSUM-based on-line covariance matrix change detector given $\tilde{\boldsymbol{C}}_i$ is given as
	\begin{equation}
		W_{i,j}(\tilde{\boldsymbol{C}}_{i-1},\tilde{\boldsymbol{C}}_i)\mathop{\lessgtr}^{H_0}_{H_1} \theta,~~\forall i,j,
		\label{CmpTheta}
	\end{equation}
	where $\theta$ is a pre-designed threshold.
	As a result, at the $i$-th covariance interval, the proposed on-line detector declares a change event at the following coherence time interval
	\begin{equation}
			j_i(\theta) = \min \mathcal{J}_i(\theta),~~\forall i,
		\label{on-lineChgDet}
	\end{equation}
	where $\mathcal{J}_i(\theta) = \big\{j~|~W_{i,j}(\tilde{\boldsymbol{C}}_{i-1},\tilde{\boldsymbol{C}}_i)> \theta\big\}$, and $\min(\mathcal{A})$ denotes the minimum element in the integer set $\mathcal{A}$.
	Moreover, the change occurrence time is estimated to be the $p_{i,j_i(\theta)}$-th coherence time interval.
	
	In the following, we explain why the change detector (\ref{CmpTheta}) is more reliable and more robust in detecting covariance change.
	In the above, we have shown that $LLR_{i,j}(\bar{\boldsymbol{h}}_{i,j},\widetilde{\boldsymbol{C}}_{i-1},\widetilde{\boldsymbol{C}}_i)>0$ may hold at some channel coherence intervals before the change point $\nu_i$, and $LLR_{i,j}(\bar{\boldsymbol{h}}_{i,j},\widetilde{\boldsymbol{C}}_{i-1},\widetilde{\boldsymbol{C}}_i)<0$ may hold at some channel coherence intervals after the change point $\nu_i$. Therefore, the LLR may oscillate around $0$ around the channel covariance matrix change point, and it is not robust to detect the change based on the LLR at a single coherence interval. Note that before the change point $\nu_i$, $LLR_{i,j}(\bar{\boldsymbol{h}}_{i,j},\widetilde{\boldsymbol{C}}_{i-1},\widetilde{\boldsymbol{C}}_i)<0$ holds for most (although not all) of $j<\nu_i$, while after the change point $\nu_i$, $LLR_{i,j}(\bar{\boldsymbol{h}}_{i,j},\widetilde{\boldsymbol{C}}_{i-1},\widetilde{\boldsymbol{C}}_i)>0$ holds for most (although not all) of $j>\nu_i$. In this case, $S(\tilde{\boldsymbol{H}}_{i,p,j},\widetilde{\boldsymbol{C}}_{i-1},\widetilde{\boldsymbol{C}}_i)$, i.e., the sum of LLRs from interval $p$ to interval $j$, can be viewed as the confidence accumulated until coherence time interval $j$ to declare a change event happened at coherence time interval $p$. (\ref{on-line_W}) indicates that $p_{i,j}$ is the estimated change point with the highest confidence, i.e., $W_{i,j}(\tilde{\boldsymbol{C}}_{i-1},\tilde{\boldsymbol{C}}_i)$. Therefore, in the change detector (\ref{CmpTheta}), no change will be declared at coherence time interval $j$ if $LLR_{i,\bar{j}}(\bar{\boldsymbol{h}}_{i,\bar{j}},\widetilde{\boldsymbol{C}}_{i-1},\widetilde{\boldsymbol{C}}_i)<0$ for many coherence time intervals $\bar{j}<j$ such that the highest confidence $W_{i,j}(\tilde{\boldsymbol{C}}_{i-1},\tilde{\boldsymbol{C}}_i)$ is smaller than a threshold, while a change is declared at coherence time interval $j$ if $LLR_{i,\bar{j}}(\bar{\boldsymbol{h}}_{i,\bar{j}},\widetilde{\boldsymbol{C}}_{i-1},\widetilde{\boldsymbol{C}}_i)>0$ for many coherence time intervals $\bar{j}<j$ such that the highest confidence $W_{i,j}(\tilde{\boldsymbol{C}}_{i-1},\tilde{\boldsymbol{C}}_i)$ is larger than a threshold.

	Note that the threshold $\theta$ plays an important role in determining the performance of the proposed on-line change detector. If $\theta$ is very large, then a change is only declared when positive LLRs are obtained over a large number of coherence time intervals such that the false alarm event rarely happens. However, in this case, the delay to detect a change, i.e., $j_i(\theta)-\nu_i$, is also very long. On the other hand, if $\theta$ is small, then the false alarm can happen easily, but the delay to detect a change is small. Therefore, in practice, $\theta$ should be carefully determined to balance the trade-off between the false alarm performance and the detection delay performance of the proposed on-line change detector.\footnote{It is worth noting that in practice, the change in the channel covariance matrix is of low frequency such that we have sufficient coherence time intervals in each covariance interval to detect any change. As a result, we ignore the missed detection events in this paper, similar to \cite{pollak1985optimal,basseville1993detection,lorden1971procedures,tartakovsky2012third,xie2021sequential}.}
	In the following, we will analytically quantify this trade-off by evaluating the effect of $\theta$ on the false alarm rate and the detection delay.
	
	\subsection{Detection Performance Analysis}\label{DetectionIndicator}
	In this subsection, we analyze the detection performance of the proposed genie-aided on-line change detectors in terms of the threshold $\theta$, which determines the trade-off between false alarm performance and detection delay performance. Particularly, similar to the conventional change detection theory \cite{pollak1985optimal}, we are interested in characterizing the long-term average performance of the on-line change detection over a sufficient large number of covariance interval. As a result, for simplicity, in the rest of this subsection, we ignore the index of the covariance interval in all the notations.
	
	First, we consider the FAR \cite{pollak1985optimal} to evaluate the long-term false alarm performance, which is defined as (given any $\theta$)
	\begin{equation}
		R_F(\theta) = \frac{1}{E_{\{\boldsymbol{H}_j,\boldsymbol{N}_j\}}\Big\{j(\theta)~\Big|j(\theta)<\nu=\infty\Big\}}.
		\label{Prob_FA}
	\end{equation}
	The denominator of (\ref{Prob_FA}) is the average time for the occurrence of a false alarm event when a change never occurs. Note that $j(\theta)$ depends on the ML channel estimation $\{\bar{\boldsymbol{h}}_j\},j=1,\dots,L_{i,1}$, which relies on the random channel $\{\boldsymbol{H}_j\}$ and noise $\{\boldsymbol{N}_j\}$, and is thus random. Hence, in (\ref{Prob_FA}), the expectation is over  $\{\boldsymbol{H}_j\}$ and $\{\boldsymbol{N}_j\}$ conditional on the event that a change never occurs, i.e., $\nu=\infty$. In practice, FAR defined in (\ref{Prob_FA}) can be interpreted as the average frequency of the false alarm events made by the proposed change detectors when a change never happens.
	
	Next, we consider the long-term delay performance. Similar to \cite{pollak1985optimal}, given any $\theta$, we are interested in the so-called CADD, which is defined as
	\begin{equation}
		T_C(\theta)= \max_{\nu\in\mathbb{Z}^+}E_{\{\boldsymbol{H}_j,\boldsymbol{N}_j\}}\Big\{j(\theta)-\nu\Big| j(\theta)\geq \nu\Big\}.
		\label{CADD}
	\end{equation}
	In (\ref{CADD}), $j(\theta)\geq\nu$ indicates that the time to declare a change should be no earlier than the given moment of change occurrence $\nu$, i.e., false alarm does not occur. Moreover, $\max_{\nu\in\mathbb{Z}^+}$ guarantees that $T_C(\theta)$ is the worst-case detection delay over all the change occurrence moments. In practice, the CADD defined in (\ref{CADD}) can be interpreted as the average detection delay of the proposed change detectors given the worst-case change moment.
	
	 It is worth noting that FAR and CADD are widely used to quantify the false alarm and detection delay performance of change detection techniques in the literature\cite{pollak1985optimal,tartakovsky2012third,xie2021sequential}. In the following, we characterize the trade-off between FAR and CADD under our proposed change detectors.
	
	\begin{thm}
		Suppose that $\text{vec}(\boldsymbol{H}_j)\sim\mathcal{CN}(\boldsymbol{0},\tilde{\boldsymbol{C}}_0)$ when $j<\nu$, $\text{vec}(\boldsymbol{H}_j)\sim\mathcal{CN}(\boldsymbol{0},\tilde{\boldsymbol{C}}_1)$ when $j\geq \nu$, where $\nu$ is the change occurrence time, and $\text{vec}(\boldsymbol{N}_j)\sim\mathcal{CN}(\boldsymbol{0},\sigma^2\boldsymbol{\boldsymbol{I}}),\forall j$. Then, under the genie-aided on-line change detector (\ref{CmpTheta}), the asymptotic trade-off between FAR and CADD is given by
		\begin{equation}
				\lim_{\theta\to\infty} \frac{T_C(\theta)}{-\log R_F(\theta)}=\frac{1}{\Phi\bigg(\big(\tilde{\boldsymbol{C}}_1+\frac{\sigma^2}{E_0}\boldsymbol{I}\big),\big(\tilde{\boldsymbol{C}}_0+\frac{\sigma^2}{E_0}\boldsymbol{I}\big)\bigg)},
			\label{Thm3}
		\end{equation}
	 where $\Phi\Big(\big(\tilde{\boldsymbol{C}}_1+\frac{\sigma^2}{E_0}\boldsymbol{I}\big),\big(\tilde{\boldsymbol{C}}_0+\frac{\sigma^2}{E_0}\boldsymbol{I}\big)\Big)$ denotes the log-determinant (LD) divergence between $\tilde{\boldsymbol{C}}_1+\frac{\sigma^2}{E_0}\boldsymbol{I}$ and $\tilde{\boldsymbol{C}}_0+\frac{\sigma^2}{E_0}\boldsymbol{I}$ given by \cite{kulis2006learning}
		\begin{equation}
			\begin{aligned}
				&\Phi\bigg(\big(\tilde{\boldsymbol{C}}_1+\frac{\sigma^2}{E_0}\boldsymbol{I}\big),\big(\tilde{\boldsymbol{C}}_0+\frac{\sigma^2}{E_0}\boldsymbol{I}\big)\bigg)\\
				=&-M-\log\bigg|\big(\tilde{\boldsymbol{C}}_1+\frac{\sigma^2}{E_0}\boldsymbol{I}\big)\big(\tilde{\boldsymbol{C}}_0+\frac{\sigma^2}{E_0}\boldsymbol{I}\big)^{-1}\bigg|\\
				&+\text{tr}\Big(\big(\tilde{\boldsymbol{C}}_1+\frac{\sigma^2}{E_0}\boldsymbol{I}\big)\big(\tilde{\boldsymbol{C}}_0+\frac{\sigma^2}{E_0}\boldsymbol{I}\big)^{-1}\Big).
			\end{aligned}
			\label{Phi}
		\end{equation}
		\label{thmon-line}
	\end{thm}
	\begin{IEEEproof}
		Please refer to Appendix A.
	\end{IEEEproof}
	
	It is worth noting that if $\tilde{\boldsymbol{C}}_1$ changes more significantly over $\tilde{\boldsymbol{C}}_0$, then $\Phi\Big(\big(\tilde{\boldsymbol{C}}_1+\frac{\sigma^2}{E_0}\boldsymbol{I}\big),\big(\tilde{\boldsymbol{C}}_0+\frac{\sigma^2}{E_0}\boldsymbol{I}\big)\Big)$ is larger. According to (\ref{Thm3}), this indicates that given the same false alarm rate, the detection delay is smaller; or equivalently, given the same detection delay, the false alarm rate is lower.

	\section{On-line Change Detection with Unknown $\{\tilde{\boldsymbol{C}}_i\}$}\label{Sec_unKnownC}
	In the previous section, to gain more insights, an on-line channel covariance matrix change detector is proposed in the ideal case where the channel covariance matrix after change is assumed to be known. However, in practice, it is usually hard to know the new channel covariance matrix in advance. In this section, we consider a more practical case when the new channel covariance after change is unknown at the beginning of each covariance interval. In the following, we will propose a WL-GLR-based covariance matrix change detector for this practical case.
	
	\subsection{WL-GLR-based Covariance Matrix Change Detector}\label{Sec_Offline_subsec1}
	The CUSUM-based on-line covariance matrix change detectors in (\ref{on-line_W}) and (\ref{CmpTheta}) for the case with known $\{\tilde{\boldsymbol{C}}_i\}$ can be extended to the case with unknown $\{\tilde{\boldsymbol{C}}_i\}$,  named generalized-likelihood-ratio-based (GLR-based) covariance change detectors, as follows.
	Because $\tilde{\boldsymbol{C}}_i$ is unknown in the change detection phase of covariance interval $i$, we need some estimation of $\tilde{\boldsymbol{C}}_i$ at each coherence time interval $j$ for change detection. Let $\hat{\boldsymbol{C}}_{i,p,j}$ denote the estimation of $\tilde{\boldsymbol{C}}_i$ at coherence time interval $j$ of covariance interval $i$, assuming that the change point is coherence time interval $p\leq j$. In practice, $\hat{\boldsymbol{C}}_{i,p,j}$ can be obtained based on all the estimated channels from coherence time interval $p$ to coherence time interval $j$, i.e., $\tilde{\boldsymbol{H}}_{i,p,j}$ (we will introduce the method to estimate $\tilde{\boldsymbol{C}}_i$ based on $\tilde{\boldsymbol{H}}_{i,p,j}$ in the following subsection). Then, with the estimated (rather than perfect) channel covariance matrix, we can modify (\ref{on-line_W}) into the following form
	\begin{equation}
		\begin{aligned}
			&\hat{p}_{i,j} = \arg\max_{1\leq p\leq j}S(\tilde{\boldsymbol{H}}_{i,p,j},\tilde{\boldsymbol{C}}_{i-1},\hat{\boldsymbol{C}}_{i,p,j}),\\
			&\widehat{W}^{(\text{GLR})}_{i,j}(\tilde{\boldsymbol{C}}_{i-1},\hat{\boldsymbol{C}}_{i,\hat{p}_{i,j},j}) = S(\tilde{\boldsymbol{H}}_{i,\hat{p}_{i,j},j},\tilde{\boldsymbol{C}}_{i-1},\hat{\boldsymbol{C}}_{i,\hat{p}_{i,j},j}),~\forall i,j,
		\end{aligned}
	\label{GLR_ele}
	\end{equation}
	where the unknown channel covariance matrix $\tilde{\boldsymbol{C}}_i$ in the sum of LLRs given in (\ref{on-line_W}) is replaced by the estimated channel covariance matrix $\hat{\boldsymbol{C}}_{i,\hat{p}_{i,j},j}$. Because $\hat{\boldsymbol{C}}_{i,\hat{p}_{i,j},j}$ may change over channel coherence intervals, there is not a recursive way to calculate $\widehat{W}^{(\text{GLR})}_{i,j}(\tilde{\boldsymbol{C}}_{i-1},\hat{\boldsymbol{C}}_{i,\hat{p}_{i,j},j})$'s as in the case with known channel covariance matrix after change \cite{Hwang10}. At last, the GLR-based on-line change detector under the case with unknown $\{\tilde{\boldsymbol{C}}_i\}$ is given by
	\begin{equation}
		\widehat{W}_{i,j}^{(\text{GLR})}(\tilde{\boldsymbol{C}}_{i-1},\hat{\boldsymbol{C}}_{i,\hat{p}_{i,j},j})\mathop{\lessgtr}^{H_0}_{H_1} \theta.
		\label{GLR}
	\end{equation}
	However, due to the estimation of $\{\tilde{\boldsymbol{C}}_i\}$, the complexity of the GLR-based change detector in (\ref{GLR_ele}) and (\ref{GLR}) is very high. Specifically, as shown in (\ref{GLR_ele}), at each coherence time interval $j$, we need to check all the possible change points from coherence time interval $p=1$ to coherence time interval $p=j$, and then calculate the corresponding covariance estimation $\hat{\boldsymbol{C}}_{i,p,j}$. In other words, until coherence time interval $j$, the new covariance matrix has been estimated by $j(j+1)/2$ times. As will be shown in the following subsection, the estimation of the covariance matrix is non-trivial. As a result, it is difficult to implement the GLR-based change detection in practice.
	
	To tackle this issue, in this paper, we adopt the WL-GLR-based change detectors. The key idea is that at each coherence time interval $j$, we only check the change point in a limited window from coherence time interval $p=j-\xi$ to coherence time interval $p=j-\bar{\xi}$ with $\xi<j$ and $\bar{\xi}<\xi$, instead of all the coherence time intervals. With the above limited window, (\ref{GLR_ele}) will be updated to
	\begin{equation}
		\begin{aligned}
			\bar{p}_{i,j}=\arg\max_{j-\xi\leq p\leq j-\bar{\xi}} S(\tilde{\boldsymbol{H}}_{i,p,j}&,\tilde{\boldsymbol{C}}_{i-1},\hat{\boldsymbol{C}}_{i,p,j}),\\
			\widehat{W}^{(\text{WL-GLR})}_{i,j}(\tilde{\boldsymbol{C}}_{i-1},\hat{\boldsymbol{C}}_{i,\bar{p}_{i,j},j}) = S&(\tilde{\boldsymbol{H}}_{i,\bar{p}_{i,j},j},\tilde{\boldsymbol{C}}_{i-1},\hat{\boldsymbol{C}}_{i,\bar{p}_{i,j},j}),\\
			&j>\xi,~\forall i.
		\end{aligned}
	\label{WL-GLR}
	\end{equation}
	Then, the WL-GLR-based change detector is given by	
	\begin{equation}
		\widehat{W}_{i,j}^{(\text{WL-GLR})}(\tilde{\boldsymbol{C}}_{i-1},\hat{\boldsymbol{C}}_{i,\bar{p}_{i,j},j}) \mathop{\lessgtr}^{H_0}_{H_1} \theta,~~j>\xi,~ \forall i.
		\label{WLGLR}
	\end{equation}
	As compared to the GLR-based change detector, under the WL-GLR-based change detector, the new channel covariance matrix is only estimated by $\xi-\bar{\xi}+1$ times at each coherence time interval. Moreover, the performance will not degrade too much due to the following reason. At each coherence time interval $j$, if the change point is very close to this coherence time interval, the number of estimated channels after change is very small. In this case, even if we calculate $\hat{\boldsymbol{C}}_{i,p,j}$ based on $\tilde{\boldsymbol{H}}_{i,p,j}$, the estimation performance is pretty poor, and it is hard to detect this change accurately. Therefore, a reliable detection is possible when the change point is not that close to the current moment. On the other hand, at each coherence time interval $j$, it is very unlikely that the change point is estimated as a coherence time interval that is far before this interval, since if this is true, then the change detector should have already detected this change in some previous moment. Due to the above reason, at each coherence time interval $j$, we can only search the changing point in a limited window $[j-\xi,j-\bar{\xi}]$, without degrading the performance too much.
	In our numerical examples in Section \ref{NumRes}, it is found that when $\xi=40$ abd $\bar{\xi}=15$, the change detection accuracy is pretty high, and the computational complexity of the algorithm is pretty low, since at each coherence interval, we just need to search $\bar{p}_{i,j}$ over at most $\xi-\bar{\xi}+1=26$ coherence intervals.

	Similar to (\ref{on-lineChgDet}) in the case of a known new covariance matrix, the WL-GLR-based change detector will declare a change detected at the following coherence time interval of covariance interval $i$
	\begin{equation}
		\widehat{j}_i(\theta) = \min\widehat{\mathcal{J}}_i(\theta),~~\forall i,
		\label{stoptime_WLGLR}
	\end{equation}
	where $\widehat{\mathcal{J}}_i(\theta)=\big\{j>\xi~\Big|~\widehat{W}_{i,j}^{(\text{WL-GLR})}(\tilde{\boldsymbol{C}}_{i-1},\hat{\boldsymbol{C}}_{i,\bar{p}_{i,j},j})> \theta\big\}$.
	
	The remaining challenge to implement the WL-GLR-based change detectors is how to calculate $\hat{\boldsymbol{C}}_{i,p,j}$ based on $\tilde{\boldsymbol{H}}_{i,p,j}$ in (\ref{WL-GLR}). In the rest of this section, we introduce an ML-based scheme to estimate the channel covariance matrix based on the method proposed in \cite{aubry2012maximum}.

	\subsection{ML Estimation of the Channel Covariance Matrix}\label{SubSec_ML}
	\vspace{-0.3cm}
	According to the change detectors (\ref{WLGLR}), based on the estimated channels from coherence time interval $p$ to coherence time interval $j$ in covariance interval $i$, i.e., $\tilde{\boldsymbol{H}}_{i,p,j}$, the ML estimation of channel covariance matrix can be obtained by solving the following problem
	\begin{equation}
		\mathcal{P}_1 \left\{
		\begin{aligned}
			\mathop{\textup{Maximize}}_{\tilde{\boldsymbol{C}}_i}&\quad S(\tilde{\boldsymbol{H}}_{i,p,j},\tilde{\boldsymbol{C}}_{i-1},\tilde{\boldsymbol{C}}_i)\\
			\textup{Subject to}&\quad \beta_l\boldsymbol{I}\preceq\tilde{\boldsymbol{C}}_i\preceq\beta_u\boldsymbol{I},
		\end{aligned}
		\right.
		\label{CostFun1_on-line}
	\end{equation}
	where $\beta_l>0$ and $\beta_u>\beta_l$ are the lower bound and upper bound for the minimum eigenvalue and the maximum eigenvalue of the estimator, respectively, to guarantee that the solution is a well-conditioned covariance matrix \cite{aubry2012maximum}. The objective function of Problem $\mathcal{P}_1$ can be re-expressed as (\ref{ObjectFunExtend}), which is shown at the top of the next page, with $\boldsymbol{S}_{i,p,j}$ being the sample covariance matrix of $\tilde{\boldsymbol{H}}_{i,p,j}$ defined as
		\begin{equation}
			\boldsymbol{S}_{i,p,j}=\frac{1}{ j-p+1}\tilde{\boldsymbol{H}}_{i,p,j}(\tilde{\boldsymbol{H}}_{i,p,j})^H.
			\label{SamCov_on-line}
		\end{equation}
	\begin{figure*}[t]
		\normalsize
		\begin{equation}
			\begin{aligned}
				S&(\tilde{\boldsymbol{H}}_{i,p,j},\tilde{\boldsymbol{C}}_{i-1},\tilde{\boldsymbol{C}}_i)=\sum_{l=p}^{j}\log\big(p(\bar{\boldsymbol{h}}_{i,l}|\tilde{\boldsymbol{C}}_i)\big)-\sum_{l=p}^{j}\log\big(p(\bar{\boldsymbol{h}}_{i,l}|\tilde{\boldsymbol{C}}_{i-1})\big)\\
				=&(j-p+1)\Big(-\log|\tilde{\boldsymbol{C}}_i+\frac{\sigma^2}{E_0}\boldsymbol{I}|-\mathrm{tr}\Big((\tilde{\boldsymbol{C}}_i+\frac{\sigma^2}{E_0}\boldsymbol{I})^{-1}\boldsymbol{S}_{i,p,j}\Big)\Big)\underbrace{-(j-p+1)M\log\pi-\sum_{l=p}^{j}\log\big(p(\bar{\boldsymbol{h}}_{i,l}|\tilde{\boldsymbol{C}}_{i-1})\big)}_{\text{\normalsize{constant terms}}},
			\end{aligned}
		\label{ObjectFunExtend}
		\end{equation}
		\hrulefill
		\vspace{-0.3cm}
	\end{figure*}

	By ignoring constant terms in the objective function in (\ref{ObjectFunExtend}), Problem $\mathcal{P}_1$ is equivalent to the following problem
	\begin{equation}
		\mathcal{P}_2 \left\{
		\begin{aligned}
			\mathop{\textup{Minimize}}_{\tilde{\boldsymbol{C}}_i}&\quad \log|\tilde{\boldsymbol{C}}_i+\frac{\sigma^2}{E_0}\boldsymbol{I}|+\mathrm{tr}\Big((\tilde{\boldsymbol{C}}_i+\frac{\sigma^2}{E_0}\boldsymbol{I})^{-1}\boldsymbol{S}_{i,p,j}\Big)\\
			\textup{Subject to}&\quad~ \beta_l\boldsymbol{I}\preceq\tilde{\boldsymbol{C}}_i\preceq\beta_u\boldsymbol{I}.
		\end{aligned}
		\right.
	\end{equation}
	
	Problem $\mathcal{P}_2$ is non-convex. However, according to \cite{aubry2012maximum}, we have the close-form solution to this problem as given in the following Theorem.
	
		\begin{thm}\label{thm_2}
		Define the eigenvalue decomposition of $\boldsymbol{S}_{i,p,j}$ as
		\begin{equation}
			\boldsymbol{S}_{i,p,j}=\boldsymbol{\Phi}^{(S)}_{i,p,j}\boldsymbol{\Lambda}^{(S)}_{i,p,j}(\boldsymbol{\Phi}^{(S)}_{i,p,j})^H,
			\label{EigenDec}
		\end{equation}
		where $\boldsymbol{\Lambda}^{(S)}_{i,p,j}=\textup{diag}(\boldsymbol{\lambda}^{(S)}_{i,p,j})$ with $\boldsymbol{\lambda}^{(S)}_{i,p,j}=[\lambda^{(S)}_{1,i,p,j},\ldots,$ $\lambda^{(S)}_{M,i,p,j}]^T$ consisting of the eigenvalues of $\boldsymbol{S}_{i,p,j}$. The elements of  $\boldsymbol{\lambda}^{(S)}_{i,p,j}$ are assumed to be sorted in ascending order, i.e., $\lambda^{(S)}_{m_s,i,p,j}\leq\lambda^{(S)}_{m_s+1,i,p,j},~\forall m_s<M$.  Moreover, $\boldsymbol{\Phi}^{(S)}_{i,p,j}$ consists of the corresponding eigenvectors of $\boldsymbol{S}_{i,p,j}$.
		Then, the optimal solution to Problem $\mathcal{P}_2$ is
		\begin{equation}
			\hat{\boldsymbol{C}}_{i,p,j}^{(\text{ML})}=  \boldsymbol{\Phi}^{(S)}_{i,p,j} (\textup{diag}(\hat{\boldsymbol{\lambda}}^{(C)}_{i,p,j}))^{-1}(\boldsymbol{\Phi}^{(S)}_{i,p,j})^H-\frac{\sigma^2}{E_0}\boldsymbol{I},~\forall i,j, \forall p<j,
			\label{ML}
		\end{equation}
		where the elements in $\hat{\boldsymbol{\lambda}}^{(C)}_{i,p,j}=[\hat{\lambda}^{(C)}_{1,i,p,j},\ldots,\hat{\lambda}^{(C)}_{M,i,p,j}]^T$ are defined by
		\begin{equation}
			\begin{aligned}
				\hat{\lambda}^{(C)}_{m,i,p,j} = \min\bigg(\max&\Big(\frac{E_0}{\beta_u E_0+\sigma^2},\frac{1}{\lambda^{(S)}_{m,i,p,j}}\Big),\frac{E_0}{\beta_l E_0 +\sigma^2}\bigg),\\
				&m=1,\dots,M.
			\end{aligned}
			\label{Optimial_lamda_on-line}
		\end{equation}
	\end{thm}

	With the ML estimator $\hat{\boldsymbol{C}}_{i,p,j}^{(\text{ML})}$ in (\ref{ML}), the proposed WL-GLR-based change detectors in (\ref{WLGLR}) will declare a change detected at the following coherence time interval of covariance interval $i$
	\begin{equation}
		\begin{aligned}
			&\widehat{j}_i^{(\text{ML})}(\theta) \\
			=& \min\big\{j>\xi\Big|\max_{j-\xi\leq p\leq j-\overline{\xi}} S(\tilde{\boldsymbol{H}}_{i,p,j},\tilde{\boldsymbol{C}}_{i-1},\hat{\boldsymbol{C}}_{i,p,j}^{(\text{ML})})\mathop{\lessgtr}^{H_0}_{H_1} \theta\big\}.
		\end{aligned}
		\end{equation}
	
	\section{Off-line Change Detector for Massive MIMO Systems}\label{SecIII}
	In the previous sections, the on-line change detectors are proposed to rapidly detect channel covariance matrix change. One notable feature of the on-line detectors is that we need to check whether a change has occurred at each coherence time interval to enable a timely detection. To achieve this goal, the new covariance matrix should be estimated by several times at each coherence time interval, even under the WL-GLR-based change detectors as shown in (\ref{WL-GLR}). However, in a massive MIMO system, the dimension of the covariance matrix is very high, and it is practically difficult to estimate the channel covariance matrix at each coherence time interval. To deal with this issue, we propose a low-complexity off-line change detection framework in this section for massive MIMO systems. Under the off-line change detection protocol, the covariance matrix change detection is merely conducted at the last coherence time interval of the first sub-interval of each covariance interval, i.e., coherence time interval $\bar{L}_1$, where $\bar{L}_1$ is carefully pre-assigned. Specifically, at covariance interval $i$, the off-line change detector is
	\begin{equation}
			\widehat{W}_i^{(\text{Off})}(\tilde{\boldsymbol{C}}_{i-1},\hat{\boldsymbol{C}}_{i,\tilde{p}_i,\bar{L}_1})= S(\tilde{\boldsymbol{H}}_{i,\tilde{p}_i,\bar{L}_1},\tilde{\boldsymbol{C}}_{i-1},\hat{\boldsymbol{C}}_{i,\tilde{p}_i,\bar{L}_1})\mathop{\lessgtr}^{H_0}_{H_1}\theta,
		\label{Offline}
	\end{equation}
	where $\hat{\boldsymbol{C}}_{i,p,\bar{L}_1}$ denotes the estimation of $\tilde{\boldsymbol{C}}_i$ according to the estimated channels from coherence time interval $p$ to coherence time interval $\bar{L}_1$, i.e., $\tilde{\boldsymbol{H}}_{i,p,\bar{L}_1}$, and $\tilde{p}_i  = \arg\max_{1\leq p\leq \bar{L}_1} S(\tilde{\boldsymbol{H}}_{i,p,\bar{L}_1},$ $\tilde{\boldsymbol{C}}_{i-1},\hat{\boldsymbol{C}}_{i,p,\bar{L}_1})$. The ML-based channel covariance matrix estimation proposed in Section \ref{SubSec_ML} can still be applied for estimating $\hat{\boldsymbol{C}}_{i,p,\bar{L}_1}$ in (\ref{Offline}). However, in a massive MIMO system, the number of available estimated channels for covariance estimation is much smaller than the dimension of the covariance matrix, i.e., $\bar{L}_1-p+1\ll M$, which will usually lead to an unstable estimation with big variance.
	In the literature, a widely used approach to generate a stable estimation when the number of samples is too small compared to the dimension of the covariance matrix is the shrinkage algorithm \cite{5484583}. Specifically, given the sample covariance matrix $\boldsymbol{S}_{i,p,\bar{L}_1}$ shown in (\ref{SamCov_on-line}), which is based on the estimated channels from coherence time interval $p$ to coherence time interval $\bar{L}_1$, the shrinkage estimator of the channel covariance matrix is given by
	\begin{equation}
		\hat{\boldsymbol{C}}^{(\text{SH})}_{i,p,\bar{L}_1}=(1-\varphi_{i,p,\bar{L}_1})\boldsymbol{S}_{i,p,\bar{L}_1}+\varphi_{i,p,\bar{L}_1}\frac{\textup{tr}(\boldsymbol{S}_{i,p,\bar{L}_1})}{M}\boldsymbol{I}-\frac{\sigma^2}{E_0}\boldsymbol{I},
		\label{shrinkage}
	\end{equation}
	where $0\leq\varphi_{i,p,\bar{L}_1}\leq1$ is the shrinkage parameter to define the weight of the sample covariance matrix in the estimation. According to \cite{5484583}, a good choice of $\varphi_{i,p,\bar{L}_1}$ is
	\begin{equation}
		\varphi_{i,p,\bar{L}_1}^* = \min\Big(\frac{-\frac{1}{M}\textup{tr}(\boldsymbol{S}_{i,p,\bar{L}_1}\boldsymbol{S}_{i,p,\bar{L}_1})+\big(\textup{tr}(\boldsymbol{S}_{i,p,\bar{L}_1})\big)^2}{\frac{\bar{L}_1-p-1}{M}\big(\textup{tr}(\boldsymbol{S}_{i,p,\bar{L}_1}\boldsymbol{S}_{i,p,\bar{L}_1})-\frac{(\textup{tr}(\boldsymbol{S}_{i,p,\bar{L}_1}))^2}{M}\big)},1\Big).
		\label{Shrin_rho}
	\end{equation}
	In this paper, we adopt the above shrinkage technique to estimate the channel covariance matrix in the massive MIMO system. Then, the off-line change detector under the shrinkage-based estimation technique is given by
\begin{equation}
		\widehat{W}_i^{(\text{Off})}(\tilde{\boldsymbol{C}}_{i-1},\hat{\boldsymbol{C}}_{i,\tilde{p}_i ,\bar{L}_1}^{(\text{SH})})= S(\tilde{\boldsymbol{H}}_{i,\tilde{p}_i,\bar{L}_1},\tilde{\boldsymbol{C}}_{i-1},\hat{\boldsymbol{C}}_{i,\tilde{p}_i ,\bar{L}_1}^{(\text{SH})})\mathop{\lessgtr}^{H_0}_{H_1}\theta,\forall i,
\end{equation}
	where $\tilde{p}_i  = \arg\max_{1\leq p\leq \bar{L}_1} S(\tilde{\boldsymbol{H}}_{i,p,\bar{L}_1},\tilde{\boldsymbol{C}}_{i-1},\hat{\boldsymbol{C}}_{i,p,\bar{L}_1}^{(\text{SH})})$.
	\begin{figure}[t]
		\vspace{-0.4cm}
		\setlength{\abovecaptionskip}{-0.1 cm}
		\centering
		\includegraphics[width=9 cm]{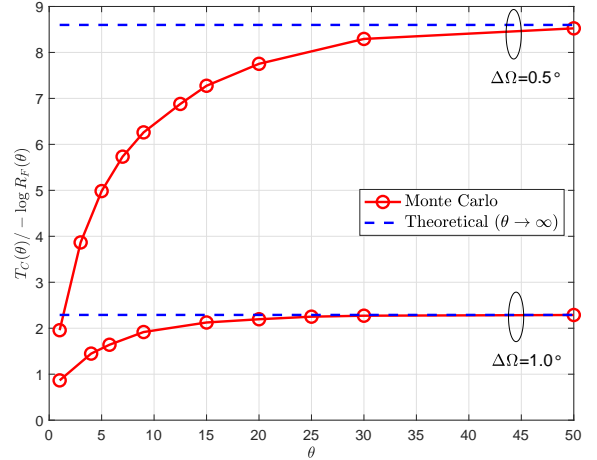}
		\caption{Verification of Theorem \ref{thmon-line} for the asymptotic ratio between CADD and FAR.}
		\label{Result1_MonteCarlonMatch}
	\end{figure}
	\begin{figure*}[t]
		\normalsize
		\begin{equation}
			\begin{aligned}
				\boldsymbol{C}_{m1,m2} = \frac{1}{2\pi}\int_{0}^{2\pi}\exp&\bigg[-j\frac{2\pi}{\alpha}D_{m1,m2}^{(T)}\sin\Omega\Big(1-\frac{\Psi^2}{4}+\frac{\Psi^2\cos(2\epsilon)}{4}\Big)+\Psi D_{m1,m2}^{(T)}\cos\Omega\sin\epsilon\\ &+D_{m1,m2}^{(R)}\sin\Omega\sin\epsilon+D_{m1,m2}^{(R)}\cos\Omega\cos\epsilon\bigg] d\epsilon, ~~ \forall m1, m2,
			\end{aligned}
			\label{ChannelMod}
		\end{equation}
		with
		\begin{subequations}
			\label{ele}
			\begin{gather}
				D^{(T)}_{m1,m2}=3\alpha\Big(\text{mod}(m1,M_t)-\text{mod}(m2,M_t)\Big),\\
				D^{(R)}_{m1,m2}=3\alpha\Big(\lfloor m1/M_t\rfloor-\lfloor m2/M_t\rfloor\Big).
			\end{gather}
		\end{subequations}
		\hrulefill
		\vspace{-0.4cm}
	\end{figure*}
	\begin{figure}[t]
		\vspace{-0.3cm}
		\setlength{\abovecaptionskip}{-0.1 cm}
		\centering
		\includegraphics[width=9 cm]{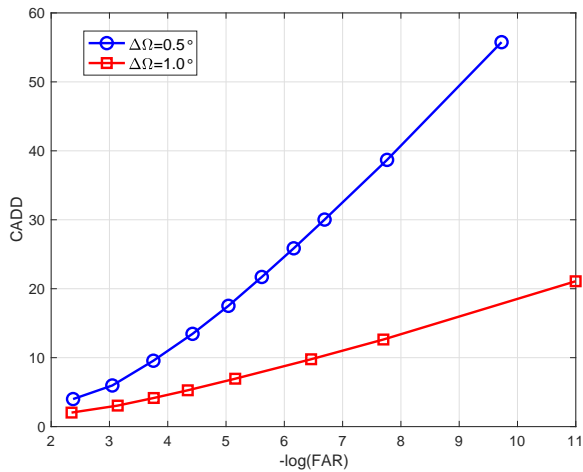}
		\caption{Performance trade-off between $-\log(\text{FAR})$ and CADD for the proposed on-line change detection with known $\{\tilde{\boldsymbol{C}}_i\}$.}
		\label{Result1_on-line}
		\vspace{-0.2cm}
	\end{figure}
	
	\section{Numerical Results}\label{NumRes}
	In this section, we present numerical results to verify the effectiveness of the proposed on-line and off-line channel covariance matrix change detection protocols. We consider both the \textit{general MIMO} and the \textit{massive MIMO} systems to demonstrate the broad applicability of the proposed change detection protocols. The path loss is modeled as $-128.1-37.6\log_{10}(d)$ in dB, where $d$ in kilometer (km) denotes the distance between the Tx and the Rx. In our numerical examples, we assume that $d=0.10$ km. Moreover, the transmit power is $23$ dBm. The bandwidth of the channel is set as $10$ MHz, and the power spectrum density of the noise is $-169$ $\text{dBm/Hz}$. In the numerical results, we use the ``one-ring" model \cite{837052} to describe the covariance matrix of the channel between the Tx and the Rx, which is shown in (\ref{ChannelMod}) at the top of this page. Particularly, $\boldsymbol{C}_{m1,m2}$ denotes the element at the $m1$-th row and $m2$-th column of the channel covariance matrix $\boldsymbol{C}$, $\Omega$ denotes the angle of departure (AoD) at the Tx side, $\Psi$ denotes the angle spread, $\theta$ denotes the angle of a possible scatter around the Rx, and $\alpha$ in meter (m) denotes the wavelength of the signal. In the numerical examples, we assume that $\Psi=30^\circ$ and $\alpha=0.15$ m, which corresponds to a carrier frequency of 2 GHz. In the rest of this section, we assume that the change in the channel covariance matrix is caused by the change of AoD, which is denoted by $\Delta\Omega$. As a result, a bigger $\Delta\Omega$ indicates that the new channel covariance matrix changes \textit{more significantly} over the covariance matrix in the previous covariance interval.
	
	\subsection{General MIMO Systems}\label{SimSec_1}
	Firstly, we illustrate the performance of the proposed on-line change detectors in general MIMO systems with moderate numbers of transmit and receive antennas. In particular, in this subsection, we assume that $M_t=8$ and $M_r=2$, and the pilot sequence length is $T=M_t=8$. Moreover, the pilot sequence $\boldsymbol{X}$ is generated based on the discrete Fourier transform (DFT) matrix.
	\subsubsection{On-line Change Detection with Known $\{\tilde{\boldsymbol{C}}_i\}$}
	First, we illustrate the performance of the proposed genie-aided on-line covariance matrix change detector (\ref{CmpTheta}) where the new channel covariance matrix after change is assumed to be known. Fig. \ref{Result1_MonteCarlonMatch} shows  the ratio between $T_C(\theta)$ and $-\log(R_F(\theta))$ when $\Delta\Omega=0.5^\circ$ and $1.0^\circ$, respectively. It is observed that as the threshold $\theta$ goes to infinity, the ratio obtained via Monte Carlo simulation asymptotically matches that from Theorem \ref{thmon-line}. It is also observed that as indicated by Theorem \ref{thmon-line}, if the change in the channel covariance matrix is more significant ($\Delta\Omega$ is larger), then the detection delay is smaller given the same false alarm rate.

	Next, Fig. \ref{Result1_on-line} shows the overall trade-off between $-\log(\text{FAR})$ and CADD of the proposed on-line change detector when $\Delta\Omega=0.5^\circ$ and $1.0^\circ$, which can be obtained by searching over all the possible values of the threshold $\theta$.
	First, it is observed that even when the AoD changes slightly by $0.5^\circ$, our proposed change detector can achieve very low CADD (delay performance) given a small FAR (reliability performance). For example, when $-\log(\text{FAR})=5$, the average detection delay is around 17 coherence time intervals.
	Moreover, the performance of the proposed on-line change detector improves significantly when the change in the channel covariance matrix is greater. For example, when $\Delta\Omega=1.0^\circ$, given $-\log(\text{FAR})=5$, the CADD is just about 6 coherence time intervals.
	
	\begin{figure}[t]
		\vspace{-0.3cm}
		\centering
		\includegraphics[width=9 cm]{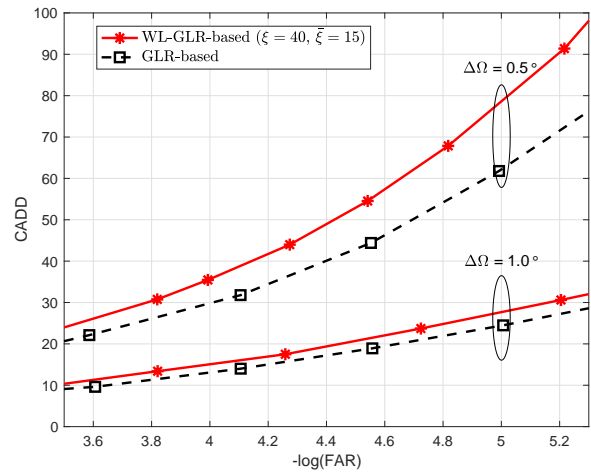}
		\vspace{-0.7cm}
		\caption{Performance trade-off between $-\log(\text{FAR})$ and CADD for the proposed on-line change detection with unknown $\{\tilde{\boldsymbol{C}}_i\}$.}
		\label{Result4_on-line}
		\vspace{-0.46cm}
	\end{figure}
	
	\subsubsection{On-line Change Detection with Unknown  $\{\tilde{\boldsymbol{C}}_i\}$}
	Next, we illustrate the performance of the proposed on-line change detector in the case with the unknown channel covariance matrix after change.
	Fig. \ref{Result4_on-line} shows the performance of the proposed WL-GLR-based covariance change detector when $\Delta\Omega=0.5^\circ, 1.0^\circ$, and $\overline{\xi}=15$. The channel covariance matrix is estimated by the ML estimator in (\ref{ML}) with $\beta_l=0.5$ and $\beta_u=4$.
	First, by comparing Fig. \ref{Result4_on-line} with Fig. \ref{Result1_on-line}, it is observed that if the new covariance matrix after change is unknown, there will be a performance loss due to the inaccurate estimation of the new covariance matrix. For example, when $\Delta\Omega=0.5^\circ$, i.e., the channel covariance matrix changes very slightly, given $-\log(\text{FAR})=4.6$, the CADD is about 60 coherence time intervals when the window size is $\xi-\bar{\xi}=40-15=25$ coherence time intervals (CADD is around 15 coherence time intervals in the ideal case with known $\{\tilde{\boldsymbol{C}}_i\}$). This indicates that a very small change can only be detected if an accurate estimation of $\{\tilde{\boldsymbol{C}}_i\}$ is available, which will take some delay in practice.
	However, when $\Delta\Omega$ increases to $1.0^\circ$ such that the change in the channel covariance matrix is more significant, given $-\log(\text{FAR})=4.6$, the CADD reduces to around 20 coherence time intervals under our proposed on-line detection scheme with the window size of 25 coherence time intervals (CADD is around 6 coherence time intervals with known $\{\tilde{\boldsymbol{C}}_i\}$). This is because if the change in the channel covariance matrix is significant, an estimation of the channel covariance matrix with moderate resolution is good enough for the change detector.
	Second, it is observed that when the window size of our proposed WL-GLR-based change detectors (\ref{WLGLR}) is 25 coherence time intervals, the performance is very close to that achieved by the GLR-based change detectors (\ref{GLR}) without a window size. However, by adding a window size, the complexity of the WL-GLR-based change detection is much reduced as shown in Table \ref{CPU_Time_WL_GLR}.
	\begin{table}
		\centering
		\caption{Average CPU time comparison between the GLR-based and WL-GLR-based change detectors when $\Delta\Omega = 1.0^\circ$.}
		\begin{tabular}{|c|c|c|c|}
			\hline
			& \multicolumn{1}{l|}{\footnotesize{-log(FAR)} } & \footnotesize{CADD} & \footnotesize{average CPU time (ms)} \\ \hline
			\begin{tabular}[c]{@{}c@{}}GLR-based \\ detector\end{tabular} &
			\multirow{2}{*}{4.5} &
			19.75 &
			18.127 \\ \cline{1-1} \cline{3-4}
			\begin{tabular}[c]{@{}c@{}}WL-GLR-based \\ detector\end{tabular} &
			&
			20.94 &
			1.143 \\ \hline
		\end{tabular}
	\label{CPU_Time_WL_GLR}
	\end{table}	
	
	\subsubsection{Comparison between On-line and Off-line Change Detectors}
\begin{figure}[t]
	\centering
	\subfigure[$\Delta\Omega=1.0^\circ$.]{
		\begin{minipage}{9cm}
			\setlength{\abovecaptionskip}{-0.5 cm}
			\includegraphics[width=9cm]{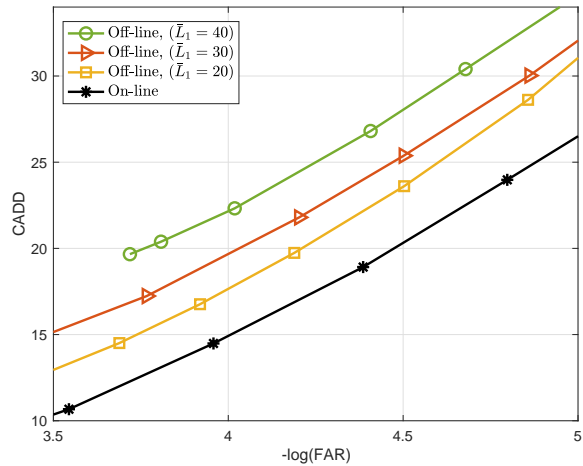}
			\label{Offline_Delta1}
		\end{minipage}
	}
	\subfigure[$\Delta\Omega=2.0^\circ$.]{
		\begin{minipage}{9cm}
			\setlength{\abovecaptionskip}{-0.1 cm}
			\includegraphics[width=9cm]{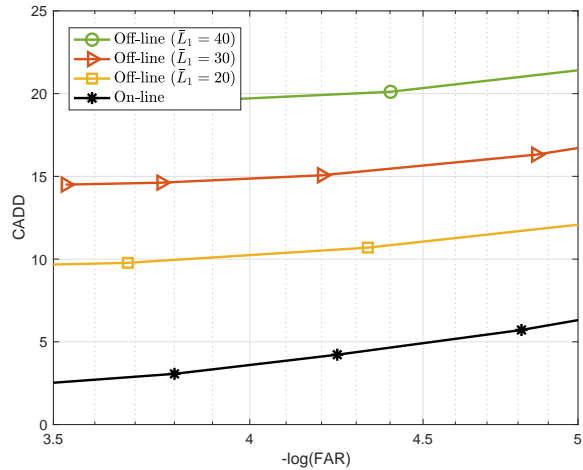}
			\label{Offline_Delta2}
		\end{minipage}
	}
	\caption{Comparison between on-line and off-line change detectors when $\Delta\Omega=1.0^\circ$ and $2.0^\circ$.}
	\label{Offline_WindowSize}
	\vspace{-0.4cm}
\end{figure}

	Next, we show the performance gain of the on-line change detectors over the off-line change detectors in the general MIMO systems. Fig. \ref{Offline_WindowSize} shows the performance comparison between the two detection schemes, where $\xi = 40$, $\bar{\xi}=15$ under the on-line scheme, and $\bar{L}_1$ is set to be $20, 30$ and $40$ under the off-line scheme. The channel covariance matrix is estimated by the ML estimator in (\ref{ML}) with $\beta_l=0.5$ and $\beta_u=4$. It is observed that given the same $-\log(\text{FAR})$, the on-line scheme achieves smaller CADD compared to the off-line scheme.
	Moreover, by comparing Fig. \ref{Offline_Delta1} and Fig. \ref{Offline_Delta2}, it is observed that when the change in the channel covariance matrix is more significant, then the gain of the on-line scheme over the off-line scheme is greater. This is because in this case, the on-line scheme can detect a change very quickly, leading to reduced detection delay.
	At last, it is worth noting that although the performance of the on-line scheme is much better than that of the off-line scheme, the complexity of the former is higher than that of the latter. As shown in Table \ref{CPU_Time_OnOff}, the off-line change detector can be obtained much faster than the on-line change detector. Such a property makes the off-line change detectors very promising in the massive MIMO systems, where it is computationally prohibitive to apply the on-line change detectors when the dimension of the channel covariance matrix is too large.
	
\begin{table}
	\centering
	\caption{Average CPU time comparison between the on-line and off-line change detectors when $\Delta\Omega=1.0^\circ$.}
	\begin{tabular}{|c|c|c|c|}
		\hline
		& \multicolumn{1}{l|}{\footnotesize{-log(FAR)} } & \footnotesize{CADD} & \footnotesize{average CPU time (ms)} \\ \hline
		\begin{tabular}[c]{@{}c@{}}On-line detector \\ ($\xi=40$)\end{tabular} &
		\multirow{2}{*}{3.67} &
		13.11 &
		0.915 \\ \cline{1-1} \cline{3-4}
		\begin{tabular}[c]{@{}c@{}}Off-line detector \\ ($\bar{L}_1=30$)\end{tabular} &
		&
		17.43 &
		0.0391 \\ \hline
	\end{tabular}
\label{CPU_Time_OnOff}
\end{table}

\begin{figure}[t]
	\vspace{-0.2cm}
	\setlength{\abovecaptionskip}{-0.1 cm}
	\centering
	\includegraphics[width=9 cm]{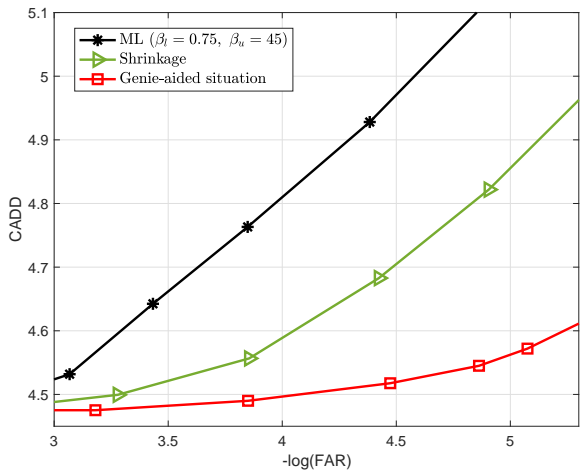}
	\caption{Off-line change detection performance in the massive MIMO system when $\Delta\Omega= 0.5^\circ$, and $\bar{L}_1=10$.}
	\label{MassiveMIMO}
	\vspace{-0.4cm}
\end{figure}
	
	\subsection{Massive MIMO Systems}
	In this subsection, we illustrate the performance of the proposed off-line change detector in massive MIMO systems. In this numerical example, we assume the channel is estimated in the uplink.
	Moreover, the Tx (user) and the Rx (BS) are equipped with $M_t=8$ and $M_r=64$ antennas, respectively.
	Therefore, the channel covariance matrix size is $M=512$. The pilot signal is generated based on DFT matrix with a dimension of 8. In this numerical example, we consider the performance of three change detection schemes: the genie-aided scheme, where the new channel covariance matrix after change is assumed to be perfectly known; the shrinkage estimation based scheme as shown in Section \ref{SecIII}; and the ML estimation based scheme as shown in Section \ref{SubSec_ML}.
	
	Fig. \ref{MassiveMIMO} shows the performance of the off-line change detection under different schemes, when $\Delta\Omega=0.5^\circ, \bar{L}_1=10$, $ \beta_l=0.75$, and $\beta_u=45$.
	It is observed that if the new covariance matrix after change is unknown, there will be a performance loss due to the imperfect estimation of the covariance matrix under both the shrinkage and the ML estimation schemes. However, it is observed that the performance loss is not significant, and the proposed off-line change detector can detect the change accurately and quickly in the massive MIMO system. For example, when $-\log(\text{FAR})=4.5$, CADD of the proposed off-line change detector under the shrinkage estimator is around 4.7 coherence time intervals.
	Moreover, it is observed that the shrinkage-based detector can achieve better performance than that achieved by the ML-based detector in the regime of $\bar{L}_1\ll M$, which indicates that the shrinkage-based estimation scheme is more powerful in estimating the large-dimension matrix.

	\section{Conclusion and Future Work}\label{Conclusion}
	This paper studied channel covariance change detection in a point-to-point MIMO communication system.
	We first considered a genie-aided on-line change detection scheme and proposed a CUSUM algorithm for the channel covariance change detection. Moreover, we analytically investigated the trade-off between the false alarm and the detection delay of the proposed on-line change detection scheme.
	Second, we considered the on-line change detection in the practical case where the channel covariance matrix after a change is unknown. To reduce the computational complexity in this case, we proposed a WL-GLR change detection algorithm where at each time instance, the new covariance matrix is estimated based on the ML technique.
	Last, we consider an off-line covariance change detection scheme for massive MIMO systems. Under this scheme, we adopted a shrinkage-based algorithm for covariance estimation.
	Our simulation results verified that the proposed on-line and off-line schemes can detect the channel covariance change with a small detection delay and a low false alarm rate.
	
	There are several interesting future research directions.
	First, it is crucial to design the change detectors when the channel does not follow the Rayleigh fading model, which may be true in the mmWave communication and the IRS-assisted communication.
	Second, it is also important to generalize our proposed scheme to the multi-user MIMO systems.
	Third, in this work, the change detection and the channel covariance matrix estimation are designed separately. However, it was shown in \cite{moustakides2012joint} that detection and estimation can be jointly designed to achieve better performance. Therefore, it is interesting to apply the technique in \cite{moustakides2012joint} for jointly detecting the change and estimating the channel covariance matrix in MIMO systems.

	\appendix
	\subsection{Proof of Theorem \ref{thmon-line}}
	We first define the worst-case average detection delay (WADD) for the proposed genie-aided on-line change detector as
	\begin{equation}
		T_W(\theta) = \max_{\nu\in\mathbb{Z}^+}\esssup \Big\{E_{\{\boldsymbol{H}_j,\boldsymbol{N_j}\}}\big\{(j(\theta)-\nu)^+\big\}\Big\},
		\label{WADD}
	\end{equation}
	where $\esssup\{\cdot\}$ denotes the essential supremum.  For any given threshold $\theta$, the WADD is lower bounded by CADD \cite{xie2021sequential}, i.e., $T_W(\theta) \geq T_C(\theta)$. Moreover, as shown in \cite{lorden1971procedures}, the asymptotic trade-off between $T_W(\theta)$ and $R_F(\theta)$ for the CUSUM-based change detector (\ref{CmpTheta}) is given by
	\begin{equation}
		\lim_{\theta\to\infty} \frac{T_W(\theta)}{-\log R_F(\theta)} = \frac{1}{\Gamma\big(\tilde{\boldsymbol{C}}_1,\tilde{\boldsymbol{C}}_0\big)},
		\label{WADDAsym}
	\end{equation}
	where $\Gamma(\tilde{\boldsymbol{C}}_1,\tilde{\boldsymbol{C}}_0)$ denotes the Kullback-Leibler (KL) divergence \cite[Eq.3.1]{kullback1997information} between the distribution $\mathcal{CN}(\boldsymbol{0},\tilde{\boldsymbol{C}}_1)$ and the distribution $\mathcal{CN}(\boldsymbol{0},\tilde{\boldsymbol{C}}_0)$, i.e.,
	\begin{equation}
		\Gamma\big(\tilde{\boldsymbol{C}}_1,\tilde{\boldsymbol{C}}_0\big)=\int p\big(\bar{\boldsymbol{h}}_{j}|\tilde{\boldsymbol{C}}_1\big)\log\frac{p\big(\bar{\boldsymbol{h}}_{j}|\tilde{\boldsymbol{C}}_1\big)}{p\big(\bar{\boldsymbol{h}}_{j}|\tilde{\boldsymbol{C}}_0\big)}d\bar{\boldsymbol{h}}_j.
		\label{Gamma}
	\end{equation}
	With $p\big({\bar{\boldsymbol{h}}_j}|\tilde{\boldsymbol{C}}_1\big)$ and $p\big({\bar{\boldsymbol{h}}_j}|\tilde{\boldsymbol{C}}_0\big)$ defined in (\ref{Prob_i}), it can be shown that $\Gamma\big(\tilde{\boldsymbol{C}}_1,\tilde{\boldsymbol{C}}_0\big)=\Phi\Big(\big(\tilde{\boldsymbol{C}}_1+\frac{\sigma^2}{E_0}\boldsymbol{I}\big),\big(\tilde{\boldsymbol{C}}_0+\frac{\sigma^2}{E_0}\boldsymbol{I}\big)\Big)$, where $\Phi\Big(\big(\tilde{\boldsymbol{C}}_1+\frac{\sigma^2}{E_0}\boldsymbol{I}\big),\big(\tilde{\boldsymbol{C}}_0+\frac{\sigma^2}{E_0}\boldsymbol{I}\big)\Big)$ is defined in (\ref{Phi}). Therefore, an upper bound for the asymptotic trade-off between the CADD $T_C(\theta)$ and the FAR $R_F(\theta)$ is given by
	\begin{equation}
		\begin{aligned}
			\lim_{\theta\to\infty}\frac{T_C(\theta)}{-\log R_F(\theta)}\leq&\lim_{\theta\to\infty}\frac{T_W(\theta)}{-\log R_F(\theta)}\\
			=&\frac{1}{\Phi\Big(\big(\tilde{\boldsymbol{C}}_1+\frac{\sigma^2}{E_0}\boldsymbol{I}\big),\big(\tilde{\boldsymbol{C}}_0+\frac{\sigma^2}{E_0}\boldsymbol{I}\big)\Big)}.
		\end{aligned}
	\end{equation}

	On the other hand, under the proposed on-line change detector, a lower bound for the asymptotic trade-off between CADD $T_C(\theta)$ and FAR $R_F(\theta)$ is given by \cite{lai1998information}
	\begin{equation}
		\lim_{\theta\to\infty}\frac{T_C(\theta)}{-\log R_F(\theta)}\geq\frac{1}{\Phi\Big(\big(\tilde{\boldsymbol{C}}_1+\frac{\sigma^2}{E_0}\boldsymbol{I}\big),\big(\tilde{\boldsymbol{C}}_0+\frac{\sigma^2}{E_0}\boldsymbol{I}\big)\Big)}.
		\label{CADDLowBound}
	\end{equation}
	Therefore, it follows that
	\begin{equation}
		\lim_{\theta\to\infty}\frac{T_C(\theta)}{-\log R_F(\theta)}=\frac{1}{\Phi\Big(\big(\tilde{\boldsymbol{C}}_1+\frac{\sigma^2}{E_0}\boldsymbol{I}\big),\big(\tilde{\boldsymbol{C}}_0+\frac{\sigma^2}{E_0}\boldsymbol{I}\big)\Big)}.
		\label{CADDAsy}
	\end{equation}

	\bibliographystyle{IEEEtran}
	\bibliography{ChgDet_New}

\begin{IEEEbiography}[{\includegraphics[width=1in,height=1.25in,clip,keepaspectratio]{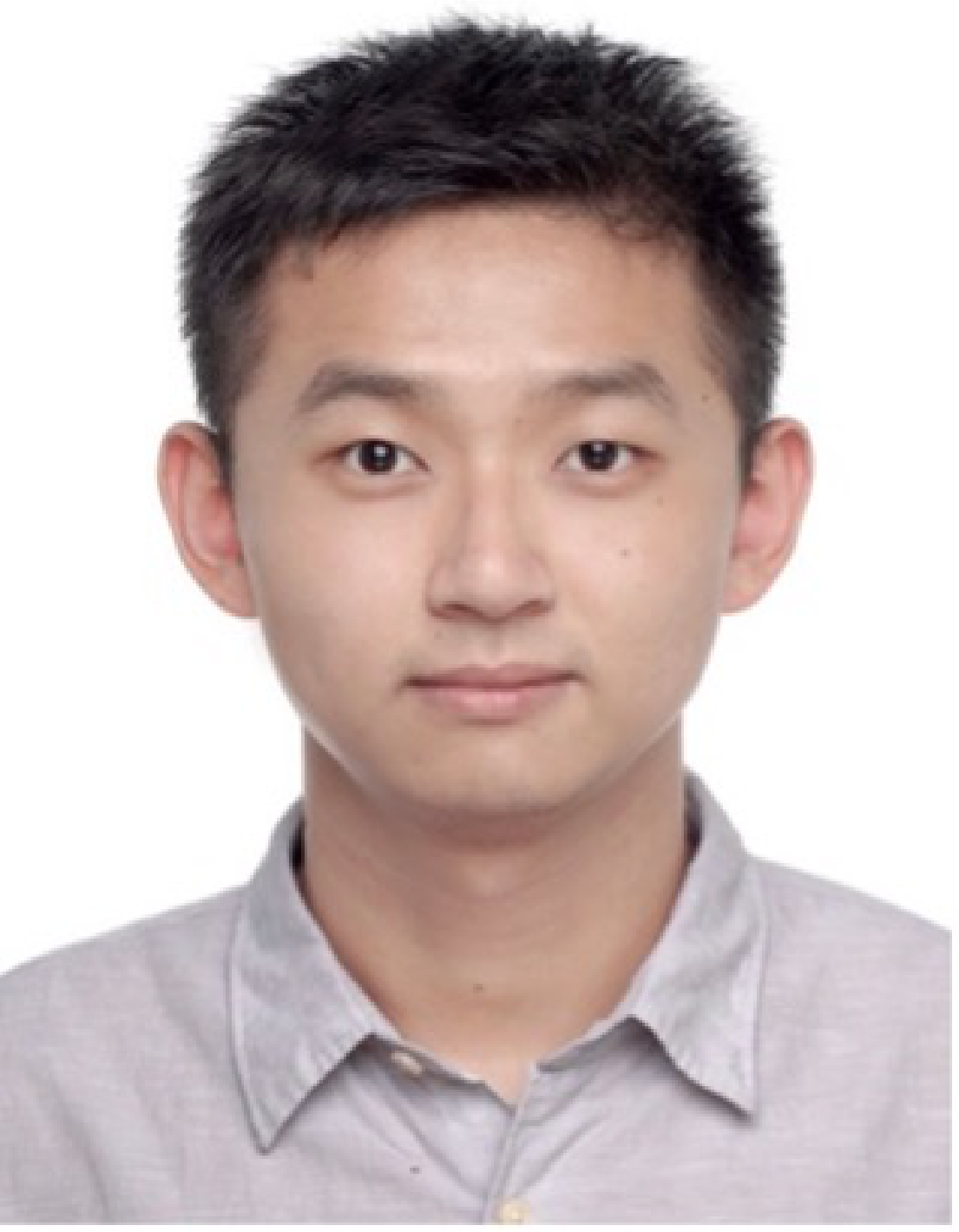}}] {Runnan Liu} (Graduate Student Member, IEEE) received his B.E. degree in communications engineering from the University of Electronic Science and Technology of China, Chengdu, China, in 2017. He is currently pursuing the Ph.D. degree with the Cooperative Medianet Innovation Center, Shanghai Jiao Tong University, Shanghai, China. Since May 2021, he has been a Visiting Researcher at the Department of Electronic and Information Engineering of The Hong Kong Polytechnic University, Hong Kong SAR, China. His research interests include the quickest change detection, MIMO, and OTFS.
\end{IEEEbiography}

\begin{IEEEbiography}[{\includegraphics[width=1in,height=1.25in,clip,keepaspectratio]{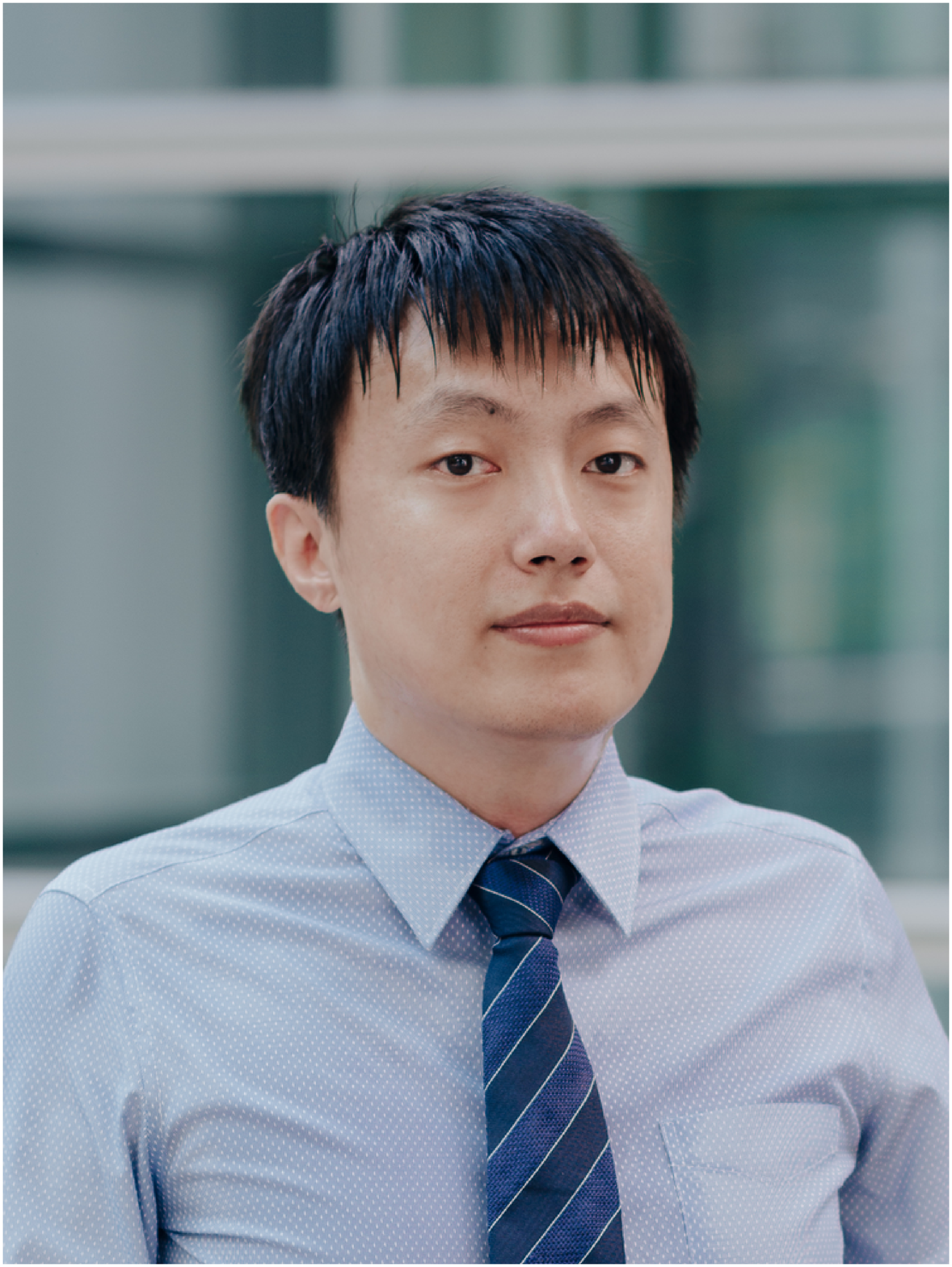}}] {Liang Liu} (Senior Member, IEEE) received the B.Eng. degree from Tianjin University, China, in 2010, and the Ph.D. degree from the National University of Singapore, Singapore, in 2014. From 2015 to 2017, he was a Post-Doctoral Fellow with the Department of Electrical and Computer Engineering, University of Toronto. From 2017 to 2018, he was a Research Fellow with the Department of Electrical and Computer Engineering, National University of Singapore. He is currently an Assistant Professor with the Department of Electronic and Information Engineering, The Hong Kong Polytechnic University. His research interests lie in the next generation cellular technologies such as machine-type communications for the Internet of Things, integrated sensing and communication, etc. He was a recipient of the 2021 IEEE Signal Processing Society Best Paper Award, the 2017 IEEE Signal Processing Society Young Author Best Paper Award, the Best Student Award of of 2022 IEEE International Conference on Acoustics, Speech, and Signal Processing (ICASSP), and the Best Paper Award of the 2011 International Conference on Wireless Communications and Signal Processing. He was recognized by Clarivate Analytics as a Highly Cited Researcher in 2018. He is an Editor of IEEE TRANSACTIONS ON WIRELESS COMMUNICATIONS. He was a Leading Guest Editor of IEEE WIRELESS COMMUNICATIONS Special Issue on ``Massive Machine-Type Communications for IoT".
\end{IEEEbiography}

\begin{IEEEbiography}[{\includegraphics[width=1in,height=1.25in,clip,keepaspectratio]{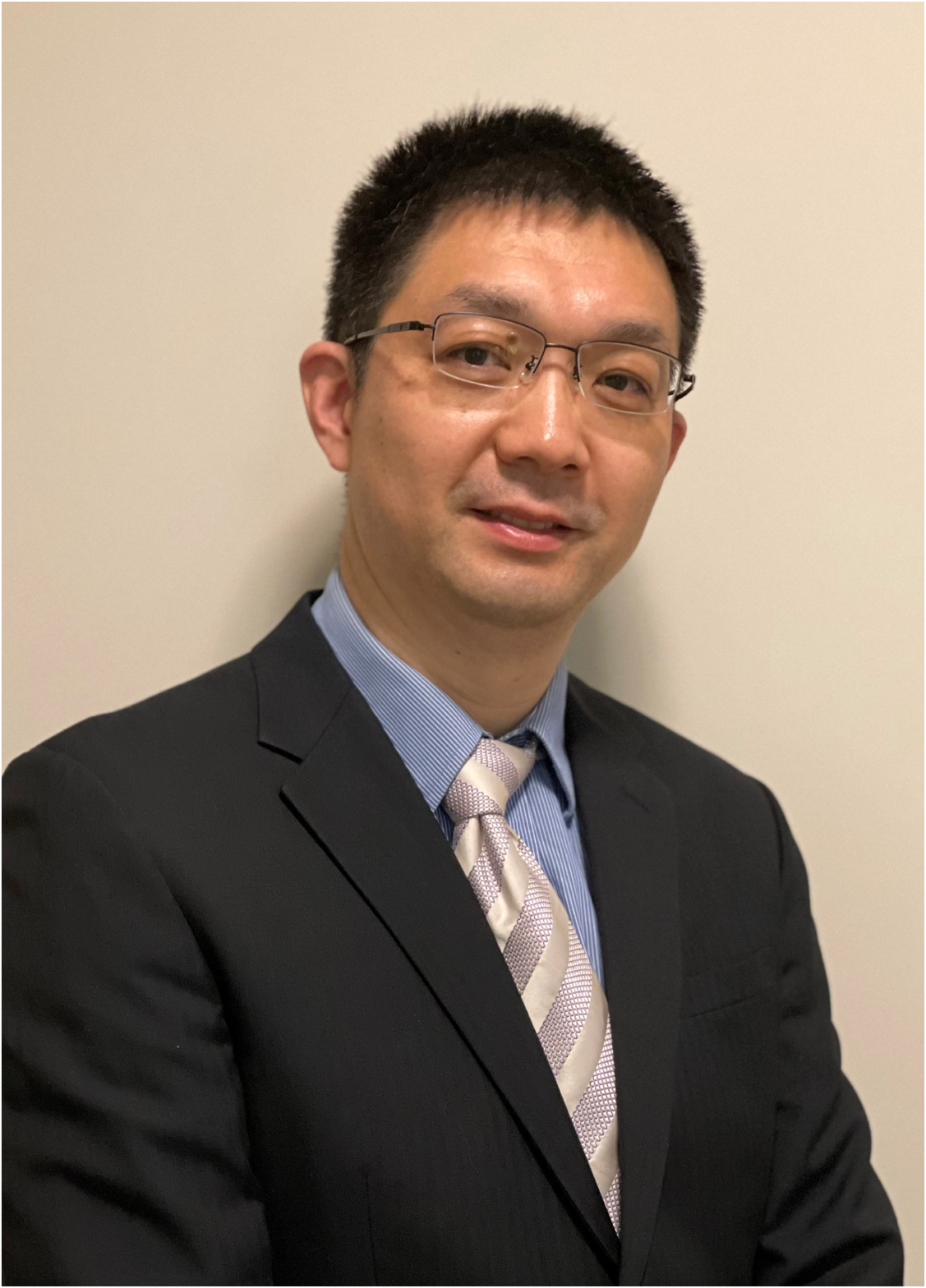}}] {Dazhi He} (Member, IEEE) is currently a Professor in the Cooperative Medianet Innovation Center of Shanghai Jiao tong University and Associate Editor of IEEE Transaction on Broadcasting. He has had in-depth research experience for the technology and standardization of terrestrial digital TV in China (DTMB), direct satellite broadcasting in China (ABS-S) and the newest global digital TV standard (ATSC3.0) . He had published more than 50 papers in the journal of IEEE and applied more than 70 patents (Including 20 PCT patents).  He ever won second prize of national scientific and technological progress award in China, first prize of the technology innovation award of National Radio and Television Administration in China, and first prize of science and technology progress award in Shanghai. His currently main research interests include 5G broadcasting, media communications, and heterogeneous network.
\end{IEEEbiography}

\begin{IEEEbiography}[{\includegraphics[width=1in,height=1.25in,clip,keepaspectratio]{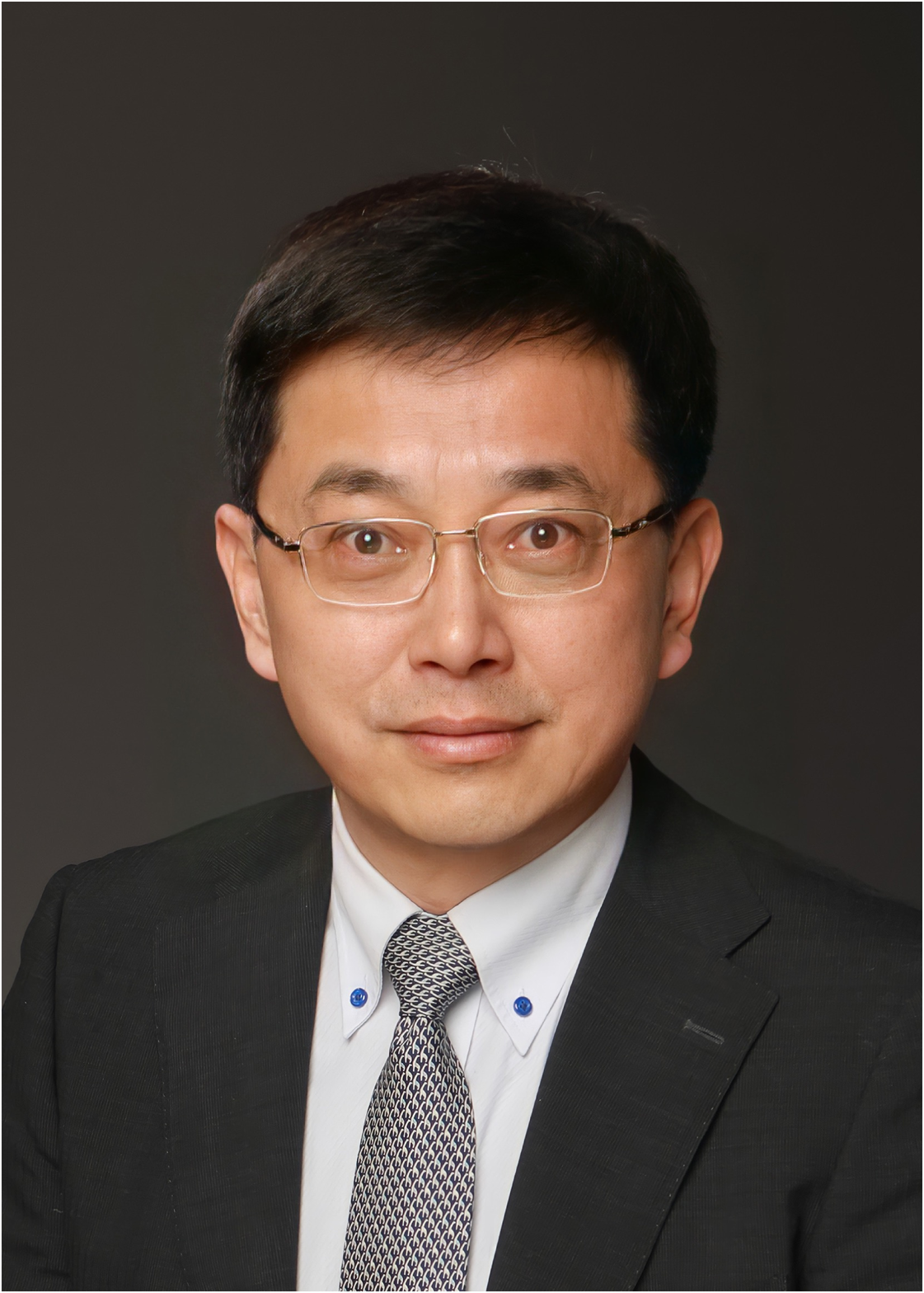}}] {Wenjun Zhang} (Fellow, IEEE) received his B.S., M.S. and Ph.D. degrees in electronic engineering from Shanghai Jiao Tong University, Shanghai, China, in 1984, 1987 and 1989, respectively. After three years' working as an engineer at Philips in Nuremberg, Germany, he went back to his Alma Mater in 1993 and became a full professor of Electronic Engineering in 1995. He was one of the main contributors of the Chinese DTTB Standard (DTMB) issued in 2006. He holds 238 patents and published more than 130 papers in international journals and conferences. He is the Chief Scientist of the Chinese Digital TV Engineering Research Centre (NERC-DTV), an industry/government consortium in DTV technology research and standardization, and the director of Cooperative Media Network Innovation Centre (CMIC), an excellence research cluster affirmed by the Chinese Government.  His main research interests include video coding and wireless transmission, multimedia semantic analysis and broadcast/broadband network convergence.
\end{IEEEbiography}

\begin{IEEEbiography}[{\includegraphics[width=1in,height=1.25in,clip,keepaspectratio]{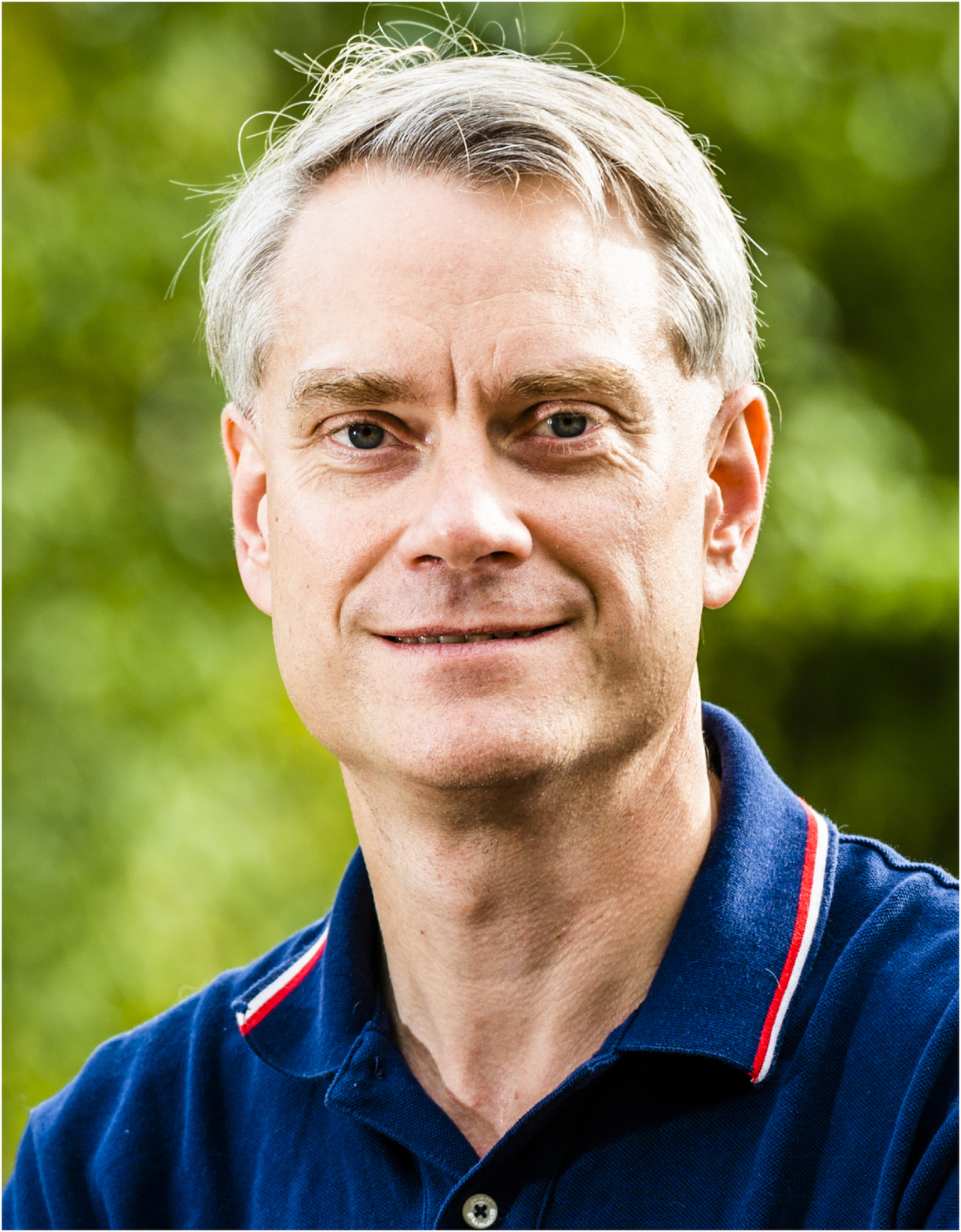}}] {Erik G. Larsson} (Fellow, IEEE) received the Ph.D. degree from Uppsala University, Uppsala, Sweden, in 2002.  He is currently Professor of Communication Systems at Link\"oping University (LiU) in Link\"oping, Sweden. He was with the KTH Royal Institute of Technology in Stockholm, Sweden, the George Washington University, USA, the University of Florida, USA, and Ericsson Research, Sweden.  His main professional interests are within the areas of wireless communications and signal processing. He  co-authored \emph{Space-Time Block Coding for  Wireless Communications} (Cambridge University Press, 2003) and \emph{Fundamentals of Massive MIMO} (Cambridge University Press, 2016).
	
He served as  chair  of the IEEE Signal Processing Society SPCOM technical committee (2015--2016), chair of  the \emph{IEEE Wireless  Communications Letters} steering committee (2014--2015), member of the  \emph{IEEE Transactions on Wireless Communications}    steering committee (2019-2022), General and Technical Chair of the Asilomar SSC conference (2015, 2012), technical co-chair of the IEEE Communication Theory Workshop (2019), and member of the IEEE Signal Processing Society Awards Board (2017--2019). He was Associate Editor for, among others, the \emph{IEEE Transactions on Communications} (2010-2014), the \emph{IEEE Transactions on Signal Processing} (2006-2010), and  the \emph{IEEE Signal  Processing Magazine} (2018-2022).
	
He received the IEEE Signal Processing Magazine Best Column Award twice, in 2012 and 2014, the IEEE ComSoc Stephen O. Rice Prize in Communications Theory in 2015, the IEEE ComSoc Leonard G. Abraham Prize in 2017, the IEEE ComSoc Best Tutorial Paper Award in 2018, and the IEEE ComSoc Fred W. Ellersick Prize in 2019.
	
\end{IEEEbiography}

\end{document}